\documentclass[pra,a4paper, amsfonts, amssymb, amsmath, reprint, showkeys, nofootinbib, twoside]{revtex4-2}

\usepackage[english]{babel}
\usepackage[utf8]{inputenc}
\usepackage[pdftex, pdftitle={Article}, pdfauthor={Author}]{hyperref} 
\usepackage{graphicx}
\usepackage{geometry}
\usepackage{scalerel,amsmath,amsthm,amssymb,comment}
\usepackage{dsfont}
\usepackage{tabto}
\usepackage{graphicx}
\usepackage{upgreek}
\usepackage{bm}
\usepackage{commath}
\usepackage{hyperref}
\usepackage{xcolor}
\usepackage{ragged2e}
\usepackage{enumitem}
\usepackage{mathtools}
\usepackage{minted}
\usepackage{float}
\usepackage{subfigure}
\usepackage{caption}
\usepackage{subcaption}

\DeclarePairedDelimiter\bra{\langle}{\rvert}
\DeclarePairedDelimiter\ket{\lvert}{\rangle}
\DeclarePairedDelimiterX\braket[2]{\langle}{\rangle}{#1\,\delimsize\vert\,\mathopen{}#2}
\makeatletter
\let\@fnsymbol@latex\@fnsymbol
\let\@fnsymbol\@alph 
\makeatother

\begin{document}
\title{Entanglement entropy of an acoustic black hole}
\author{P. C. van de Graaf and H. T. C. Stoof}
\affiliation{Institute for Theoretical Physics and Center for Extreme Matter and Emergent Phenomena, \\
Utrecht University, Princetonplein5, 3584 CC Utrecht, The Netherlands. }
\date{\today}

\begin{abstract}
We introduce a method to numerically compute the entanglement entropy of an acoustic black hole. It is shown that the entanglement entropy of sufficiently large subregions scales linearly with size and thus shows a volume law instead of an area law. The origin of this scaling can be traced back to the non-separable long-distance correlations due to the production of phonon pairs at the horizon. The system is shown to be locally thermal, such that the part of the entanglement entropy scaling with volume is well approximated by the thermal entropy of the outgoing Hawking radiation. 

\end{abstract}

\maketitle

\section{Introduction}
It is well known that black holes bear many similarities with thermodynamical systems. The groundwork for this equivalence has been laid out in the 1970's by Bekenstein, Hawking and others \cite{Bekenstein,Hawking}. In particular, it was found that black holes have a natural entropy that scales only with the area of the event horizon. This discovery has further led to the ideas of holography \cite{Susskind_holo, hooft2009dimensionalreductionquantumgravity} and the black-hole information paradox \cite{Susskind:2006oza}. However, the nature of the corresponding microstates has remained elusive so far. 

When it was discovered that the entanglement entropy of quantum fields satisfy similar area laws, it has been suggested that the entropy of the black hole could possibly be interpreted as the entanglement entropy due to the hidden quantum fields behind the horizon \cite{Bombelli}. The actual computation to explore this idea further proves to be very difficult, however, and is in many cases ill defined due to ultraviolet divergences near the horizon. One possible solution has been put forward by 't Hooft in the form of a so-called brick-wall model, which introduces a frequency cutoff to remove the divergences \cite{tHooft}. The choice of cutoff is unfortunately somewhat unsatisfactory, as it is taken such to reproduce the desired area law but does not allow for more insight into the underlying physical mechanisms that should overcome this problem. It thus appears that a complete computation would require knowledge of the presently unknown physics of gravity at the Planck scale. Several of such computations have therefore been done in the context of string theory \cite{Strominger_1996}, and holography \cite{Ryu_2006}.

The aim of this work is to show that analogue black holes can also provide a promising alternative. First proposed by Unruh in 1981, these are experimentally realizable condensed-matter systems that can mimic the behavior of field theories in curved spacetime \cite{Unruh}. In contrast to gravity, these systems possess a fully specified microscopic description, with well defined sub-Planckian physics, which naturally remove the near-horizon divergences. Various analogue black holes have been proposed in different media, such as those in surface waves \cite{Water}, superfluid helium \cite{Jacobson_1998}, light in dispersive media \cite{light_dispersive_media}, trapped-ion rings \cite{Horstmann_2010}, ultracold fermions \cite{Giovanazzi_2005}, light in non-linear liquids \cite{light}, Weyl semi-metals \cite{Volovik_weyl_fermions}, exciton-polariton condensates \cite{Solnyshkov_2011,Nguyen_2015}, and magnons \cite{artim, Magnons}. In this work, we consider a superfluid Bose-Einstein condensate, in which Hawking radiation has been observed in 2014 \cite{Jeff2}. 

The acoustic black hole in the latter system is required to be infinite in size, as reflective boundary conditions causes instabilities due to repeated amplification at the horizon \cite{Fischer,Steinhauer}. Moreover, infrared and ultraviolet regulators are required to precisely control divergences in the entanglement entropy. We show that a good candidate, which can satisfy all these conditions with slight modifications, is the analogue black hole first proposed by Recati {\it et al.\ }\cite{Recati_2009}. In our work, the scaling of the entanglement entropy of the acoustic black hole is studied using the correlator method \cite{Symplectic}. This approach has an advantage with respect to the more standard brute-force integration technique introduced by Bombelli and Srednicki \cite{Bombelli,srednicki}, since it allows the computation of the entanglement entropy for an arbitrary Gaussian state. This is necessary, since the black-hole state is highly dependent on the chosen boundary conditions. We ultimately find that transsonic configurations of the black hole introduce long-distance correlations, which heavily influence the entanglement entropy between subregions. In the presence of the acoustic black-hole horizon, this leads to a volume law, instead of the usual logarithmic scaling of a conformal field theory. It is also shown that this volume-law scaling is well described as the local thermal entropy of the outgoing Hawking radiation.

The paper is built up as follows. In Sec.\ II we introduce the acoustic black-hole setup as proposed in Ref.~\cite{Recati_2009}. The required modifications to the theory in order to reliably compute the entanglement entropy are then discussed in Sec.\ III. In Sec.\ IV, the different scattering solutions for the quantum fluctuations on top of the acoustic black hole are constructed. From these results, the Hawking spectrum of outgoing radiation can be computed using the resulting $S$ matrix of the scattering problem. Moreover, we can present in Sec.\ V the structure of long-distance correlations in the natural vacuum of the theory. Finally, Sec.\ VI discusses the computation of the entanglement entropy of the acoustic black hole. The scaling with the size of the subregion is first discussed in the homogeneous Bogoliubov theory without superflow as a benchmark. Then we turn to the topic of most interest and study the influence of the long-distance correlations in the black-hole background on the entanglement entropy, after which the general structure for the size dependence of the entanglement entropy far from the horizon can be proposed. We end by presenting our conclusions in Sec.\ VII. 

\begin{figure}
    \centering
    \includegraphics[width=0.8\linewidth]{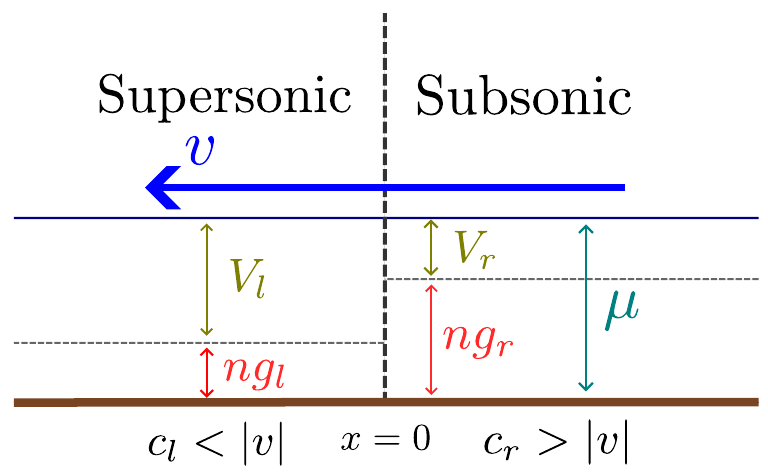}
    \caption{A sketch of the setup of an acoustic black hole in a Bose-Einstein condensate. The external potentials are chosen such that the chemical potential and density are constant. Here the $V_{l/r}$ and $g_{l/r}$ denote the external potentials and coupling strengths, respectively, where the index $l/r$ denotes the left or right region from $x=0$. The coupling strengths $g_{l/r}$ are chosen such that the velocity profile is transsonic, i.e., $c_l < |v|$, and $c_r > |v|$, where $c_{l/r} = \sqrt{g_{l/r}n/m}$ are the local speeds of sound and $v$ is the superfluid velocity.}
    \label{fig:setup} 
\end{figure}

\begin{figure*}[t]
    \centering
    \includegraphics[width=0.8\textwidth]{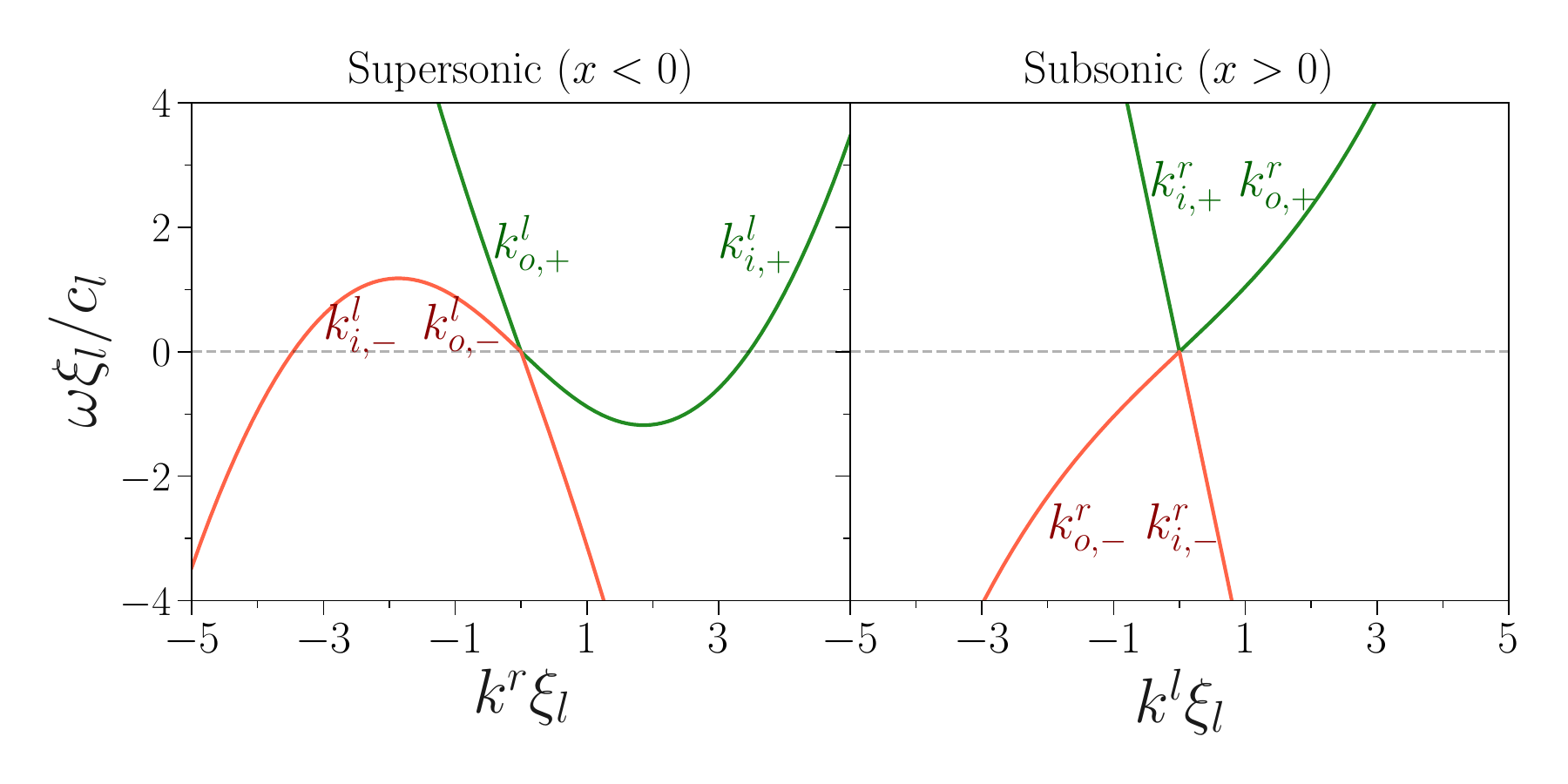}
    \caption{The dispersion relation of the continuous Bogoliubov theory in the supersonic (left), and the subsonic (right) regions.  Positive and negative normed branches are colored green and red respectively. The momenta are labeled by $i/o$ referring to ingoing and outgoing modes respectively, and $\pm$ to differentiate between positive and negative normed solutions. The flow profile has ${c_l}/{|v|} = {1}/{2}$ and ${c_r}/{|v|}= {3}/{2}$.}
    \label{fig:dispersions}
\end{figure*}

\section{Acoustic black-hole setup}
For completeness, we first give a short review of the acoustic black hole proposed by Recati {\it et al.\ }\cite{Recati_2009}. It consists of two separate one-dimensional homogeneous Bose-Einstein condensates that are connected at the horizon at $x=0$, and Fig.~\ref{fig:setup} presents a schematic illustration of the setup. Each of the condensates behaves as a superfluid, and is described by the Gross-Pitaevski equation for the macroscopic wave function $\phi_0(x,t)$, we use units such that $\hbar = 1$ throughout, 
\begin{align}
i\partial_t \phi_0 = -\frac{1}{2m}\partial_x^2 \phi_0 + g|\phi_0|^2 \phi_0 + V\phi_0.
\end{align}
Here $g$ is the coupling strength, which is related to the interatomic scattering length $a$ via $g = 4\pi a/m$, and $V$ is the external potential \cite{Henk_boek}. The potential and the coupling strength are chosen to have constant but different values in both regions $x>0$ and $x<0$. In order to ensure a constant density $n$ of the condensate, the external potential is set such that the chemical potential is constant everywhere, $\mu = \mu_l = \mu_r$, which leads to the condition $\mu = g_ln  + V_l= g_rn + V_r$. The full solution is then given by the plane wave
\begin{align}
\phi_0(x,t) = \sqrt{n}e^{i(kx-\omega t)}.
\end{align}
This corresponds to a flowing condensate with a constant background velocity $v = k/m$. 

Gaussian fluctuations in the condensate are described by the Heisenberg equation
\begin{align}
\left[i(\partial_t + v\partial_x) + \frac{\xi c}{2}\partial_x^2 - \frac{c}{\xi} \right]\hat{\phi} = \frac{c}{\xi}\hat{\phi}^\dagger, \label{bdg}
\end{align}
with the local healing length $\xi = 1/\sqrt{gmn}$ and local speed of sound $c=\sqrt{gn/m}$ \cite{Henk_boek}. They are related via $\xi_l c_l = \xi_r c_r = 1/m$. The healing length $\xi$ acts as the natural Planck scale of the theory. At length scales larger than the healing length, the dynamics of the Bogoliubov theory can be shown to resemble that of a massless scalar field in a curved spacetime \cite{Unruh}. At smaller length scales, a quadratic dispersion of ordinary free particles is recovered. The left-hand side of the system is chosen to have $c_l > |v|$, and the right-hand side $c_r < |v|$, such that the flow transitions from subsonic to supersonic at the horizon. 

The field operator $\hat{\phi}(x,t)$ can be written in terms of annihilation and creation operators by the mode-function decomposition
\begin{align}
\hat{\phi}(x,t) = \sum_j \int_0^\infty d\omega \left[u_{\omega}^j(x,t)\hat{a}_{\omega,j} + {v^{j}_{\omega}}^*(x,t)\hat{a}^\dagger_{\omega,j}\right]. \label{decomposition}
\end{align}
Note that the sum is over all independent mode functions at the same frequency and the integral is only over positive frequencies. The latter is due to the time-reversal symmetry of the Heisenberg equation, which makes sure that the sum over the normal and anomalous parts of the expansion automatically takes into account both positive and negative-frequency solutions. For the fields to satisfy the correct commutation relations, the mode functions have to obey the normalization condition
\begin{align}
\int dx &\left[ u_{\omega}^j(x){u_{\omega'}^{j'}}^*(x) -  v_{\omega}^j(x){v^{j'}_{\omega'}}^*(x) \right]
= \delta(\omega - \omega')\delta^{jj'}. \label{norm:1}
\end{align}
Moreover, the equations of motion for the mode functions follows from Eq.~(\ref{bdg}) and are given by the Bogoliubov-de Gennes equations
\begin{align}
\left[i(\partial_t + v \partial_x) + \frac{\xi c}{2}\partial_x^2 - \frac{c}{\xi} \right]u_{\omega}^j(x,t) &= \frac{c}{\xi}v_{\omega}^j(x,t), \nonumber\\
\left[-i(\partial_t + v \partial_x) + \frac{\xi c}{2}\partial_x^2 - \frac{c}{\xi} \right]v_{\omega}^j(x,t) &= \frac{c}{\xi}u_{\omega}^j(x,t), \label{eom2:2}
\end{align}
where $c/\xi$ is implicitly understood to be different in the left and right regions of the acoustic black hole. Note that the solutions satisfy time-reversal symmetry, which interchanges the mode functions as
\begin{align}
u_{\omega}^j(x,t) &\to {v_{\omega}^j}^*(x,t), \nonumber\\
v_{\omega}^j(x,t) &\to {u_{\omega}^j}^*(x,t). \label{symm}
\end{align}
Clearly, only one of these solutions can be normalized according to Eq.~(\ref{norm:1}) and the other has a norm with opposite sign.

We solve the full inhomogeneous problem by first substituting the plane-wave {\it ansatz} $u_k(x,t) = U_k e^{ikx - i\omega t}$ and $v_k(x,t) = V_k e^{ikx - i\omega t}$ in the two homogeneous regions. Then Eq.~\eqref{eom2:2} leads to the condition
\begin{align}
\frac{V_k}{U_k} = \frac{\xi}{c}\left(\omega - vk\right) - \frac{\xi^2}{2}k^2 -1 \equiv R_k,
\end{align}
together with the dispersion relation
\begin{align}
\left(\omega - vk\right)^2 = c^2k^2 + \frac{c^2\xi^2k^4}{4}. \label{disp}
\end{align}
The dispersions for the subsonic and supersonic regions are shown in Fig.~\ref{fig:dispersions}. 
Note that the normalization condition in Eq.~(\ref{norm:1}) for the individual plane-wave modes can only be satisfied when $|U_k|>|V_k|$. Modes that violate this condition are said to have negative norm, which for the supersonic system correspond to the modes with momenta $k^l_{i,-}$ and $k^l_{o,-}$. They have a probability flux in opposite direction to their group velocity. The presence of these two plane-wave modes in the supersonic region will ultimately lead to negative-normed states for the full inhomogeneous problem. This issue can be accounted for by making use of the symmetry in Eq.~(\ref{symm}) or, equivalently, by interchanging the corresponding creation and annihilation operators. The normalization of the plane-wave modes in the homogeneous system is given by
\begin{align}
U_k &= \left|2\pi\left(1-{R_k}^{2}\right)\left(\frac{d k}{d\omega}\right)^{-1}\right|^{-\frac{1}{2}}, \\
V_k &= \left|2\pi\left(1-R_k^{-2}\right)\left(\frac{d k}{d\omega}\right)^{-1}\right|^{-\frac{1}{2}}. \label{U/V}
\end{align}
This normalization ensures that the modes have a unit flux. 

The general solution of the sought for inhomogeneous mode functions are a superposition of these plane waves, i.e., 
\begin{align}
u_{\omega}^j(x,t) = \begin{cases}\sum_k A_{k}^{j,l} U_k^{l} e^{ik^{l}x - i\omega t} \quad\text{if $x\leq0$}, \\\sum_k A_{k}^{j,r} U_k^{r} e^{ik^{r}x - i\omega t} \quad \text{if $x>0$}.\end{cases} \nonumber\\
v_{\omega}^j(x,t) = \begin{cases}\sum_{k} A_{k}^{j,l} V_k^{l} e^{ik^{l}x - i\omega t} \quad\text{if $x\leq0$},\\\sum_{k} A_{k}^{j,r} V_k^{r} e^{ik^{r}x - i\omega t} \quad \text{if $x>0$}.\end{cases}\label{modes}
\end{align}
The sum is over all roots of the dispersion relation in Eq. (\ref{disp}) for a certain frequency $\omega$. Conservation of probability at the horizon requires that the sum of the flux of the incoming modes equals the sum of the flux of the outgoing modes, so for each value of $j$ we have
\begin{align}
\sum_{k_{i},l/r} \pm |A_{k_{i}}^{l/r}|^2 = \sum_{k_{o},l/r} \pm |A_{k_{o}}^{l/r}|^2= 1,
\end{align}
where the $\pm$ sign depends on whether the mode with momentum $k$ has a positive or negative norm. 
The amplitudes $A^{l/r}$ are determined by matching the mode functions across the horizon. This is done by demanding that they are continuous and differentiable over $x=0$, which leads to the matching conditions on the amplitudes
\begin{align}
W_l
\begin{pmatrix}
A^l_{k_1}\\
A^l_{k_2}\\
A^l_{k_3}\\
A^l_{k_4}
\end{pmatrix} = W_r\begin{pmatrix}
A^r_{k_1}\\
A^r_{k_2}\\
A^r_{k_3}\\
A^r_{k_4}
\end{pmatrix},
\end{align}
where
\begin{align}
W^{l/r} = 
\begin{pmatrix}
U^{l/r}_{k_1} & U^{l/r}_{k_2} &  U^{l/r}_{k_3}&  U^{l/r}_{k_4}\\
ik_1U^{l/r}_{k_1} & ik_2U^{l/r}_{k_2} &  ik_3U^{l/r}_{k_3}&  ik_4U^{l/r}_{k_4}\\
V^{l/r}_{k_1} & V^{l/r}_{k_2} &  V^{l/r}_{k_3}&  V^{l/r}_{k_4}\\
ik_1V^{l/r}_{k_1} & ik_2V^{l/r}_{k_2} &  ik_3V^{l/r}_{k_3}&  ik_4V^{l/r}_{k_4}
\end{pmatrix} . \nonumber
\end{align}
In order to complete the mode decomposition, a basis $\{ j \}$ of solutions to these scattering equations needs to be defined at each frequency. 
\begin{figure*}[t]
    \centering
    \includegraphics[width=0.8\textwidth]{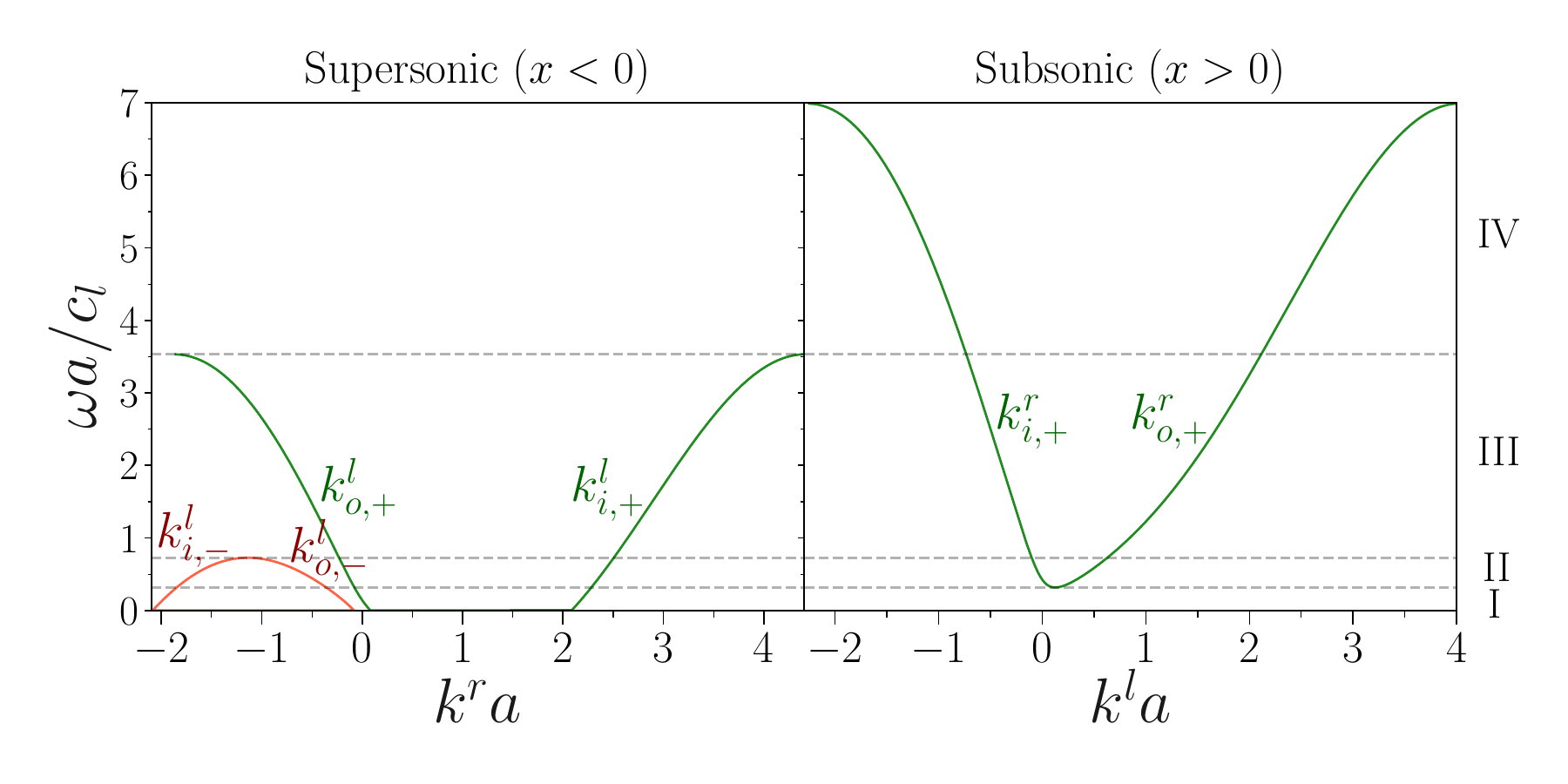}
    \caption{The dispersion relation of the modified Bogoliubov theory in the supersonic (left), and the subsonic (right) regions for positive frequencies. Positive and negative normed branches are colored green and red, respectively. The momenta are labeled by $i,o$ referring to ingoing and outgoing modes respectively, and $\pm$ to differentiate between positive and negative normed solutions. Different matching regimes are labeled by the roman numerals I to IV. The flow profile has ${c_l}/{|v|} = {1}/{2}$ and ${c_r}/{|v|}= {3}/{2}$. For visibility, a relatively large mass parameter ${\omega_0a}/{c_l} = 0.001$ and small healing length $\xi_l = 0.1a$ are used. }
    \label{fig:dispersion}
\end{figure*}

\section{Regularization}
There are several problems that arise when computing the entanglement entropy from the theory as described above. It is well known that the entanglement entropy in field theories is generally an ultraviolet divergent quantity \cite{Bombelli}. It is therefore necessary to introduce a regulator to ensure finiteness. One possible approach is to introduce a cutoff length scale $a$ in the theory, by restricting the integral in Eq.~(\ref{decomposition}) to frequencies corresponding to incoming modes with wavelength of at most $a$. Due to the fact that this condition leads to different cut-off frequencies for the left and right regions of the acoustic black hole, we have found that this leads to several complications, including the violation of the Heisenberg uncertainty conditions for Gaussian states, which is highly problematic when computing the entanglement entropy. 

It is therefore more accurate to introduce an infinite tight-binding model with lattice spacing $a$. For the matching conditions at the horizon to be consistent, the derivatives are required to include at most nearest-neighbor hopping terms
\begin{align}
\delta_x f(x) &= \frac{f(x+a) - f(x-a)}{2a},\\
\delta_x^2 f(x) &= \frac{f(x+a) - 2f(x) + f(x-a)}{a^2},
\end{align}
such that the eigenvectors are still given by plane waves $\phi_k = e^{ikx}$, but with eigenvalues
\begin{align}
\frac{\delta_x \phi_k}{\phi_k} &= \frac{i}{a}\sin\left(ak\right), \\
\frac{\delta_x^2 \phi_k}{\phi_k} &= -\frac{4}{a^2}\sin^2\left(\frac{ak}{2}\right), 
\end{align}
instead of the continuum results $ik$ and $-k^2$  respectively, obtained in the limit $a \rightarrow 0$. 

Apart from the ultraviolet divergences, there is also the issue of zero modes. Since the theory is not gapped, the zero mode leads to a divergence in the entanglement entropy, as it carries correlations over infinite distances. The most common and straightforward approach to regularize this is to add a mass-like term $\omega_0$ to the dispersion, and then take the limit $\omega_0 \to 0$. This method of regularization works well for small system sizes, as the main features of the entanglement entropy will then be shown to be independent of this gap. However, for the black-hole system, which is required to be infinite in size, it is not as straightforward to study the $\omega_0 \to 0$ limit. Nonetheless, from general arguments, it will still be possible to draw several conclusions as we discuss later in more detail. Another possibility would be to consider higher-dimensional systems, which naturally remove the infrared divergence. We leave this idea open for now and do not explore it further here. 

With the introduced gap and tight-binding model, the Heisenberg equation in Eq.~(\ref{bdg}) transforms into
\begin{align}
\left[i(\partial_t + v\delta_x) + \frac{\xi c}{2}\delta_x^2 - \frac{c}{\xi} + \omega_0\right]\hat{\phi} = \frac{c}{\xi}\hat{\phi}^\dagger,
\end{align}
where the normalization of the mode functions in Eq.~(\ref{norm:1}) is now given by
\begin{align}
a \sum_{n=-\infty}^\infty &\left[ u_{\omega}^j(na){u_{\omega'}^{j'}}^*(na) -  v_{\omega}^j(na){v^{j'}_{\omega'}}^*(na) \right]\nonumber \\
&= \delta(\omega - \omega')\delta^{jj'}. \label{norm_aangepast}
\end{align}
This leads to the same plane-wave normalization as in Eqs.~(10) and (\ref{U/V}), but with
\begin{align}
R_k = & \frac{\xi}{c}\left(\omega - \frac{v}{a}\sin\left(ka\right)\right) \nonumber \\ &- \frac{2\xi^2}{a^2}\sin^2\left(\frac{ka}{2}\right) - \frac{\omega_0\xi}{c}-1.
\end{align}
Finally, the dispersion relation is transformed into
\begin{align}
\left(\omega - \frac{v}{a}\sin\left(ka\right)\right)^2& = \frac{4c^2+4\xi c\omega_0}{a^2}\sin^2\left(\frac{ka}{2}\right) \nonumber\\+ \frac{4c^2\xi^2}{a^4}&\sin^4\left(\frac{ka}{2}\right) + \omega_0^2 + \frac{2c \omega_0}{\xi}. \label{dispd}
\end{align}
In Fig. \ref{fig:dispersion} we show this dispersion relation for both the supersonic and subsonic regions. It is similar to that of the continuous theory, except that it is gapped in the subsonic region, and both regions have now a natural frequency cutoff $\omega_{\text{m}}^l$ and $\omega_{\text{m}}^r$. 

The matching conditions are modified to the discrete theory by demanding that the left and right modes have equal values at $x=0$ and $x=a$, such that u$_l(0) = u_r(0)$ and $u_l(a) = u_r(a)$. The nearest-neighbor hopping terms then ensure that the matched equations form full solutions globally. This leads to the matching matrices
\begin{align}
W^{l/r} = 
\begin{pmatrix}
U^{l/r}_{k_1} & U^{l/r}_{k_2} &  U^{l/r}_{k_3}&  U^{l/r}_{k_4}\\
U^{l/r}_{k_1}e^{iak_1} & U^{l/r}_{k_2}e^{iak_2} &  U^{l/r}_{k_3}e^{iak_3}&  U^{l/r}_{k_4}e^{iak_4}\\
V^{l/r}_{k_1} & V^{l/r}_{k_2} &  V^{l/r}_{k_3}&  V^{l/r}_{k_4}\\
V^{l/r}_{k_1}e^{iak_1} & V^{l/r}_{k_2}e^{iak_2} &  V^{l/r}_{k_3}e^{iak_3}&  V^{l/r}_{k_4}e^{iak_4}
\end{pmatrix}.\nonumber
\end{align}

\begin{figure*}[t]
    \centering
    \includegraphics[width=0.45\textwidth]{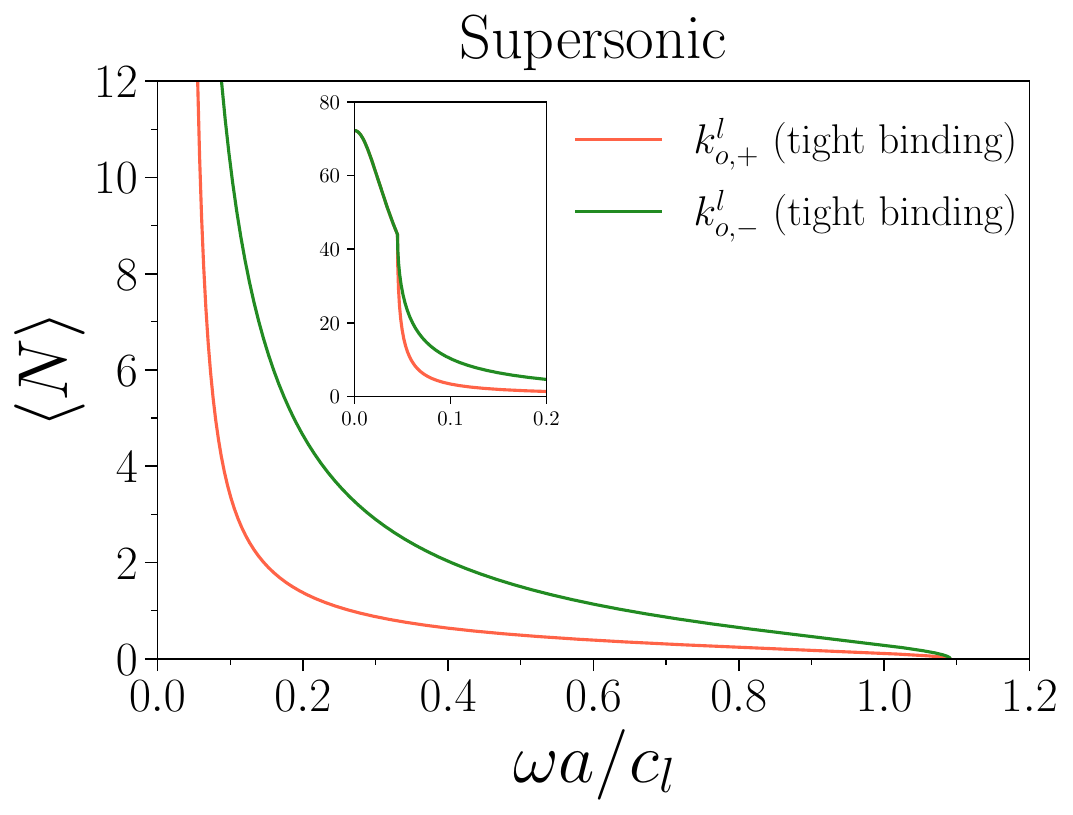}
    \includegraphics[width=0.45\textwidth]{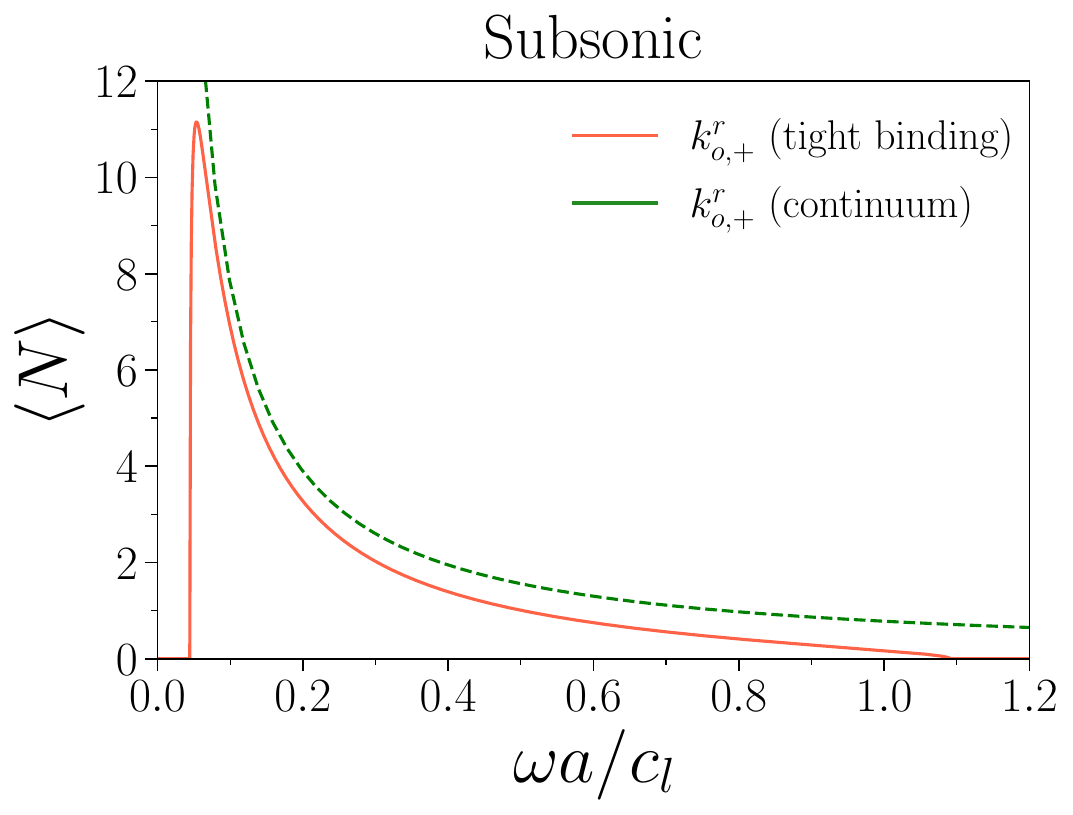}
    \caption{The occupation numbers of outgoing Hawking pairs in the supersonic (left) and subsonic (right) regions as a function of frequency for the tight-binding model with gap ${\omega_0a}/{c_l} = 5\cdot 10^{-5}$, and healing length $\xi_l = 4a$. The flow profile is chosen such that ${c_l}/{|v|} = {1}/{2}$ and ${c_r}/{|v|}={3}/{2}$. The subsonic region is compared to the analytical results for the continuous theory found by Ref.~\cite{Recati_2009}. }
    \label{fig:flux}
\end{figure*}

\section{Hawking Radiation}
The above matching procedure to obtain the solutions to the Bogliubov-de Gennes equation for our piecewise-homogeneous system can be seen as solving a (multi-channel) scattering problem and the results can thus be described using the scattering matrix $S$ that relates the amplitudes for the ingoing and outgoing modes. The vacuum $|{\it in}\rangle$ of the system is naturally defined by the absence of incoming particles at $t=-\infty$. This condition corresponds to a choice of basis consisting of only the ingoing scattering solutions, such that the amplitudes $A^{l/r}_{i,k}$ are all proportional to the $S$ matrix elements. There are three different ingoing channels, corresponding to the momenta $k^l_{i,-}$, $k^l_{i,+}$, and $k^r_{i,+}$. The scattering matrix will be labeled by upper indices for the ingoing mode, and lower indices for the outgoing mode, such that the coefficient relating the momentum $k_{i,-}^{l}$ with momentum $k^r_{o,+}$ is denoted by $S^{l,-}_{r,+}$, and so forth. 

All amplitudes require a separate matching at every frequency. To do so most conveniently, the frequency range is partitioned into four physically different regimes. Regime I is given by the frequencies below the mass gap $\omega<\omega_{\text{gap}}$ in the subsonic region. The solutions are here given by full reflections of the incoming $k^l_{i,-}$ and $k^l_{i,+}$ modes off the horizon. Regime II describes the energy range where there is hydrodynamical behavior, $\omega_{\text{gap}}<\omega < \omega_{h}$, where $\omega_h$ is the maximum frequency for the negative-normed modes, separating regime II and III. In this regime, all incoming modes contribute to the dynamics, and the dispersion is close to linear for the modes $k^r_{o,+}/k^l_{o,-}$ and $k^r_{i,+}/k^l_{o,+}$, with group velocities $v_g = v + c$ and $v_g = v-c$, respectively. It is this regime that is most relevant in the Hawking process and in the long-distance correlations, as will be discussed in the following sections. In regime III, $\omega_{h} < \omega < \omega^l_{\text{m}}$, only positive-normed states $k^l_{i,+}$ and $k^r_{i,+}$ contribute. The sub-Planckian physics start to dominate and the dispersions becomes quadratic. In the final regime, regime IV, there are no more propagating modes in the supersonic region, such that there is only a full reflection of the incoming mode from the subsonic region off the horizon, similar to regime I. 

In the appendix, the specifics for the scattering conditions can be found. The computations are done numerically. Analogous to the basis of ingoing modes, a basis of outgoing modes can be constructed. They correspond to channels with outgoing momenta $k^l_{o,-}$, $k^l_{o,+}$ and $k^r_{o,+}$. An observer far from the horizon will naturally define its vacuum state using a mix of the ingoing and outgoing modes. Its observed particle content is computed from the Bogoliubov transformation between the two bases. 

In regime I, the mode functions are related via
\begin{align}
\begin{pmatrix} 
u_{in,\omega}^{l,-}\\
u_{in,\omega}^{l,+}\\
\end{pmatrix} = \begin{pmatrix}
S^{l,-}_{l,-} & S^{l,-}_{l,+} \\
S^{l,+}_{l,-} & S^{l,+}_{l,+} \\
\end{pmatrix}
\begin{pmatrix} 
u_{out,\omega}^{l,-}\\
u_{out,\omega}^{l,+}\\
\end{pmatrix}.
\end{align}
The first mode has a negative norm, and the second has a positive norm. This leads to mixing between the creation and annihilation operators
\begin{align}
\begin{pmatrix} 
\hat{a}_{\omega,l,-}^{out\dagger}\\
\hat{a}_{\omega,l,+}^{out}\\
\end{pmatrix} = \begin{pmatrix}
S^{l,-}_{l,-} & S^{l,+}_{l,-} \\
S^{l,-}_{l,+} & S^{l,+}_{l,+} \\
\end{pmatrix}
\begin{pmatrix} 
\hat{a}_{\omega,l,-}^{in\dagger}\\
\hat{a}_{\omega,l,+}^{in}\\
\end{pmatrix}.
\end{align}
The occupation numbers of outgoing particles that an observer far from the horizon measures, are found from $\bra{in}\hat{a}^{out\dagger}_{\omega,j}\hat{a}_{\omega',j}^{out}\ket{in} \equiv N_j(\omega) \delta(\omega - \omega')$ and are thus given here by
\begin{align}
N_{l,-} &= |S^{l,+}_{l,-}|^2, \nonumber\\
N_{l,+} &= |S^{l,-}_{l,+}|^2 \label{occupation1}.
\end{align}

In regime II, we analogously obtain that
\begin{align}
\begin{pmatrix} 
\hat{a}_{\omega,l,-}^{out\dagger}\\
\hat{a}_{\omega,l,+}^{out}\\
\hat{a}_{\omega,r,+}^{out}
\end{pmatrix} = \begin{pmatrix}
S^{l,-}_{l,-} & S_{l,-}^{l,+} & S^{r,+}_{l,-}\\
S^{l,-}_{l,+} & S_{l,+}^{l,+} & S_{l,+}^{r,+}\\
S^{l,-}_{r,+} & S^{l,+}_{r,+} & S_{r,+}^{r,+}
\end{pmatrix}
\begin{pmatrix} 
\hat{a}_{\omega,l,-}^{in\dagger}\\
\hat{a}_{\omega,l,+}^{in}\\
\hat{a}_{\omega,r,+}^{in}
\end{pmatrix},
\end{align}
such that the outgoing occupation numbers of phonons are given by
\begin{align}
N_{l,-} &= |S^{l,+}_{l,-}|^2 + |S^{r,+}_{l,-}|^2, \nonumber\\
N_{l,+} &= |S^{l,-}_{l,+}|^2, \nonumber\\
N_{r,+} &= |S^{l,-}_{r,+}|^2. \label{occupation2}
\end{align}
Since only negative-normed states can lead to mixing, there is no particle creation in regimes III and IV. 

In Fig. \ref{fig:flux} we plot the occupation number of the different outgoing phonons. It should be noted that the spectrum is not exactly thermal. This is a result of the sub-Planckian physics that is introduced due to the stepwise velocity profile. Moreover, the maximum frequency for the negative normed modes ensures the occupation is zero above the frequency $\omega_h$. If instead a smooth profile is chosen which varies at length scales larger than the healing length, a thermal spectrum should be recovered in the long-wavelength limit, with a temperature proportional to the surface gravity, i.e., $T = {\kappa}/{2\pi}$. Note also that due to the introduction of a gap there is a natural cutoff to the occupation numbers at low frequencies. This is in stark contrast to the gapless theory, where the occupation number diverges. In the subsonic region this leads to an occupation of zero at low frequencies. In the supersonic region, the maximum value is capped off, as can be seen from the inset in Fig. \ref{fig:flux}.

\section{Correlations}
It is essential now to introduce the real (density) field $\rho = \phi + \phi^\dagger$ and its conjugate momentum $\pi = {i}(\phi^\dagger - \phi)/2$. The equal time two-point correlation functions can be computed from the mode decomposition in the {\it in} basis as
\begin{align}
&\langle \rho(x)\rho(y)\rangle_{in}  = \sum_{j\in\{l+,r\}} \int_0^{\omega_{\text{m}}^r} d\omega \left[u_{\omega}^j(x) + v_{\omega}^j(x)\right] \nonumber\\
& \times \left[{u_{\omega}^j}^*(y) + {v_{\omega}^j}^*(y)\right] + \int_0^{\omega^r_{m}} d\omega \left[{u_{\omega}^{l,-}}^*(x) + {v_{\omega}^{l,-}}^*(x)\right] \nonumber\\
& \times \left[u_{\omega}^{l,-}(y) + v_{\omega}^{l,-}(y)\right], \label{corr_real}
\end{align}
where the mode functions $u_{\omega}^j$ and $v_{\omega}^j$ are as in Eq.~(\ref{modes}), and $\omega_{\text{m}}^r$ is the maximum frequency for modes. The other two-point correlators have a similar expression, which can be found in the appendix. The computation of the correlators is done using numerical integration. 

\begin{figure}
    \centering
    \includegraphics[width=1\linewidth]{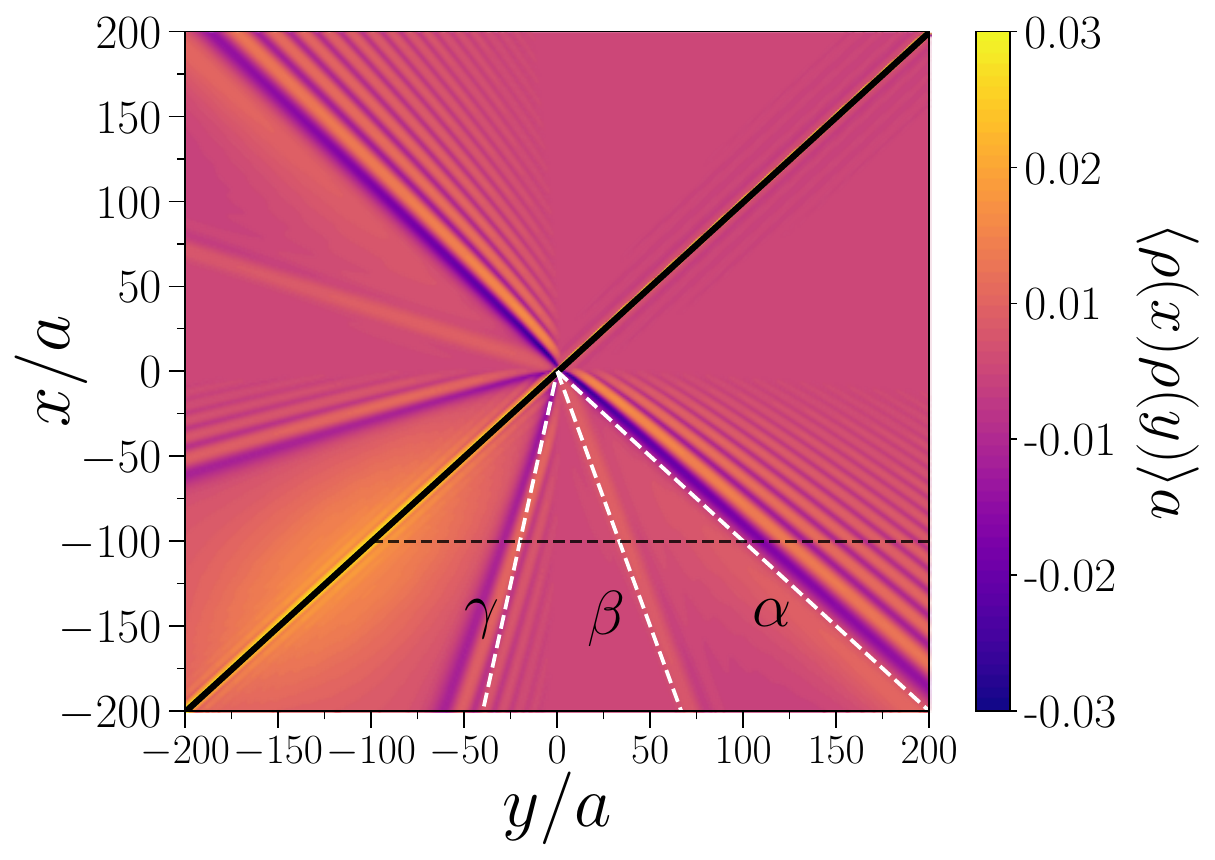}
    \caption{The density-density correlation function of the analogue black hole. The flow profile has ${c_l}/{|v|} = {1}/{2}$ and ${c_r}/{|v|}= {3}/{2}$, with the healing length $\xi_l = {a}$ and gap ${\omega_0a}/{c_l}=10^{-4}$. In order to see the long-distance correlations clearly, the short-distance correlations (diagonal) are masked. Dashed white lines correspond to the peaks as described by Eq.~(\ref{group_vels}), and are labeled by the Greek letters $\alpha$, $\beta$ and $\gamma$. The horizontal dashed line denotes the cut corresponding to Fig.~\ref{fig:doorsnee}. }
    \label{fig:correlations}
\end{figure}

In Fig.~\ref{fig:correlations} we show the resulting $\rho-\rho$ correlation function. The off-diagonal features signify long-distance correlations in this system. This is due to the Hawking pairs that are produced at the horizon \cite{Recati_2009, Carusotto_numerical}. When such a pair is produced, the phonons propagate outward at their corresponding group velocity $v_g = v\pm c$. This results in peaks along the directions
\begin{align}
x= \frac{v + c_r}{v+c_l}y, \quad x= \frac{v+c_r}{v-c_l}y, \quad  x = \frac{v+c_l}{v-c_l}y. \label{group_vels}
\end{align}
Note that if the gap size is increased, the group velocity is no longer constant, which results in quantum diffusion and a widening of these features. If the long-distance correlations are not separable, there is entanglement over large distances across the horizon. This will be responsible for the volume-law scaling in the entanglement entropy. The long-distance correlations only appear for transsonic velocity profiles, since the main contribution comes from the incoming mode $k_{i,-}^l$, which is absent in subsonic flow.

\section{Entanglement Entropy}
There are several methods for computing the entanglement entropy in field theories, the most common of which was developed by Bombelli and Srednicki \cite{Bombelli,srednicki}. It starts from the Hamiltonian, and constructs the vacuum as the state of minimal energy, after which the degrees of freedom outside the subregion are integrated out. However, for a black-hole system, this computation has been shown to raise problems. The presence of the supersonic region introduces degrees of freedom with harmonic potentials having the wrong sign, hence leading to instabilities. Despite this issue, several computations have been done for the near-horizon entanglement entropy, by only considering subsystems with a hard wall outside the black hole \cite{EE1,EE2,EE3}. Such a method, however, fails to incorporate the contribution due to Hawking radiation, which will significantly modify the structure of the entanglement entropy. In our specific system, the instabilities manifests itself in the form of the negative-normed states at positive frequencies. As discussed earlier, they effectively carry a negative energy, which means that a global vacuum cannot be defined properly, as there is no lower bound to the energy. In fact, if reflective or periodic boundary conditions are introduced, these unstable modes are amplified, which results in imaginary eigenvalues of the Bogoliubov-de Gennes equations \cite{Fischer,Steinhauer} via the mechanism of symmetry breaking in pseudo-Hermitian matrices \cite{symm_breaking}. In this case, backreaction effects need to be introduced, which considerably complicate the computations. This is why it is crucial for the analogue black-hole system to be infinite in size.

The {\it in} and {\it out} vacua as described in the previous section are in fact excited states of the system, which require a different approach to compute their entanglement entropy. One such alternative is the so-called correlator method, which can be naturally generalized to work for any arbitrary Gaussian state as shown by Casini and Huerta \cite{Casini&Huerta}. The approach is based on the fact that, for a Gaussian problem, all locally observable quantities in any subsystem of size $N$ can be computed directly from the symmetrized correlation matrix $\Gamma_{ii'} = \langle r_i r_{i'} + r_{i'} r_i \rangle$, where the $r_i$ are given by $r_i = \rho_i$ if $i = 1, \dots, N$, and $r_{i} = \pi_{i-N}$ if $i = N+1, \dots, 2N$ \cite{Symplectic}. As a result, the entanglement entropy can be fully determined from the two-point correlation functions restricted to the subsystem of interest. 

In fact, it can be shown that the eigenvalues of $\Gamma\Omega$, where 
\begin{equation} 
\Omega = \begin{pmatrix}
0&\mathbb{I} \\
-\mathbb{I} & 0
\end{pmatrix} 
\end{equation} 
and $\mathbb{I}$ denotes the $N \times N$ unit matrix, are invariant under symplectic, i.e., canonical, transformations. These eigenvalues are referred to as the symplectic eigenvalues, and come in pairs of opposite sign $i\sigma_1,..,i\sigma_N$ and $-i\sigma_1,...,-i\sigma_N$. Williamson's theorem states that there exists a symplectic transformation to a system of uncoupled harmonic oscillators, such that the correlation matrix becomes diagonal $\Gamma \ = \text{diag}(\lambda_1,...,\lambda_{2N})$ \cite{Williamson}. By relating the latter eigenvalues to the known symplectic eigenvalues, the entanglement entropy can be computed to be
\begin{align}
S_E =
\sum_{i=1}^N & \left[ \left(\sigma_i + \frac{1}{2}\right)\log\left(\sigma_i + \frac{1}{2}\right) \right. \nonumber \\ & ~~ - \left. \left(\sigma_i - \frac{1}{2}\right)\log\left(\sigma_i - \frac{1}{2}\right) \right].
\end{align}
This method allows numerical computation of the entanglement entropy of any partition that divides the system into a subsystem consisting of a finite number of points $A$, and a subsystem $B$, such that the Hilbert space obeys $\mathcal{H} = \mathcal{H}_A \otimes \mathcal{H}_B$ using only the local correlators on subsystem $A$. 

\subsection{Homogeneous Bogoliubov theory}
\begin{figure}[t]
    \centering
    \includegraphics[width=0.93\linewidth]{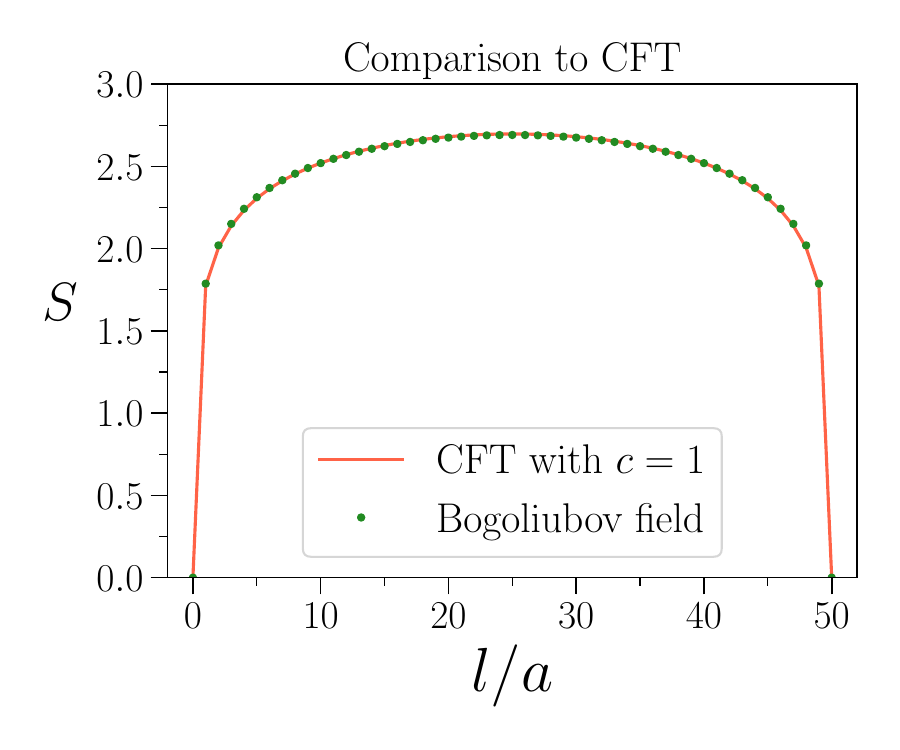}
    \caption{The entanglement entropy of a finite-size homogeneous Bogoliubov theory for $\xi = 0.01a$ and gap $\omega_0 a/c_r=10^{-8}$, compared to the exact results from conformal field theory. }
    \label{fig:conformal}
\end{figure}

To the best of our knowledge, the structure of the entanglement entropy for the homogeneous Bogoliubov theory without superflow has not been studied explicitly before. Moreover, it is an important test case for us to understand before we embark on the calculation for the acoustic black hole. The length scales that then parametrize the problem are given by the healing length $\xi_r$, the lattice step size $a$, the inverse gap $c_r/\omega_0$, and the system size $L$. First consider the situation where the system size is small compared to the length scale corresponding to the gap, i.e., $L \ll c_r/\omega_0$. In this limit, the scaling of the entanglement entropy is independent on the size of the gap $\omega_0$. This ensures that its behavior is fully determined by the ratio between the healing length and the step size. In the limit when $\xi \ll a$, the theory is conformal with central charge $c=1$. It is known that the scaling of the entanglement entropy of a conformal theory of $N = L/a$ lattice sites with periodic boundary conditions is given by \cite{Calabrese_2009}
\begin{align}
S_E = \frac{c}{3}\log \left(\frac{N}{\pi}\sin\left(\frac{\pi n}{N}\right)\right) + c',
\end{align}
where $n$ is the number of sites of the subregion $A$ not integrated out and $c'$ is a non-universal constant. Fig.~\ref{fig:conformal} shows that the Bogoliubov theory satisfies this scaling in the limit ${\xi}/{a} \ll 1$. For small subregions $n \ll N$, the entanglement entropy satisfies a logarithmic scaling $S_E = \log(n)/3 + c'$, which for a conformal field theory in one dimension is usually referred to as the area law. 

\begin{figure}[t]
    \centering
    \includegraphics[width=1\linewidth]{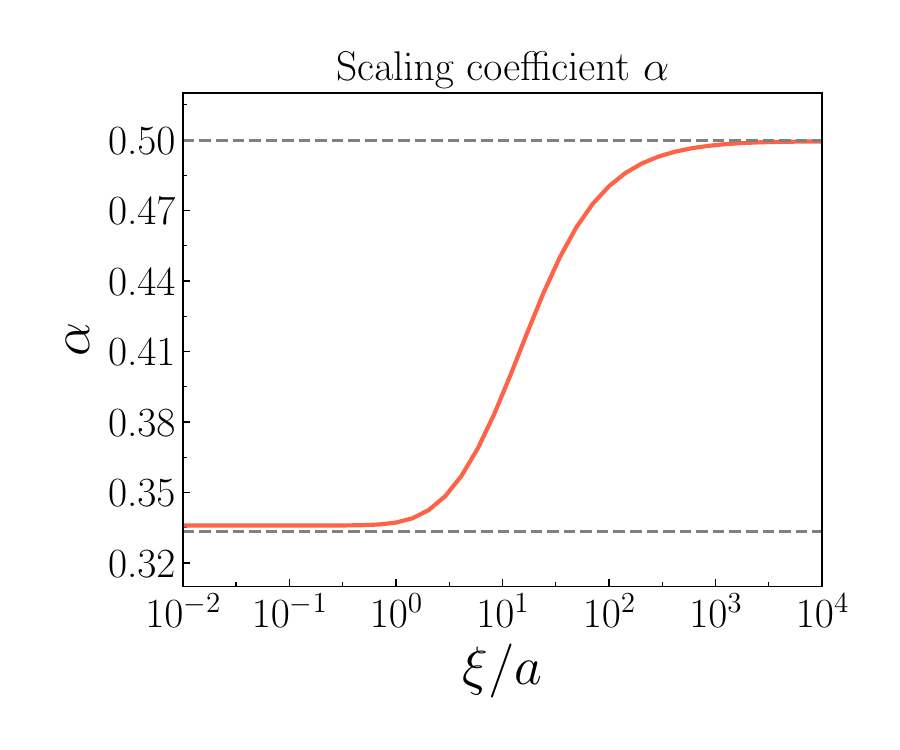}
    \caption{The logarithmic scaling coefficient $\alpha$ of the homogeneous Bogoliubov theory as a function of the relative length scale ${\xi}/{a}$ in the limit when $l\ll L \ll c_r/\omega_0$. For low values of ${\xi}/{a}$ it converges to the known conformal-field-theory value of ${1}/{3}$ and for high values to ${1}/{2}$.}
    \label{fig:scaling}
\end{figure}

For intermediate values of ${\xi}/{a}$, there are no known analytical results. The scaling appears, however, to remain logarithmic at short length scales, but with a varying prefactor. This leads us to introduce the following {\it ansatz} for the entanglement entropy of the homogeneous Bogoliubov theory
\begin{align}
S_E = \alpha \left(\frac{\xi}{a} \right) \log\left(\frac{l}{a}\right) + c',
\end{align}
where the prefactor $\alpha$ is a function of the system parameters $\xi/a$ and $l \equiv n a$ is the subsystem size. It can be determined by fitting this relation to the computed entanglement entropy. Fig.~\ref{fig:scaling} shows the behavior of this prefactor as a function of ${\xi}/{a}$. As expected, at low ${\xi}/{a}$, the scaling tends to the value $\alpha = {1}/{3}$ known from conformal field theory. At higher values of ${\xi}/{a}$, the prefactor seems to converge to a value of $\alpha = {1}/{2}$. When the mass term is varied, the logarithmic scaling remains unaffected at short length scales. All the mass dependence is absorbed into the non-universal constant factor $c'$. This validates the use of a small mass parameter as a cutoff, as long as only system sizes where $L \ll c_r/\omega_0$ are considered. 

\begin{figure}[t]
    \centering
    \includegraphics[width=1\linewidth]{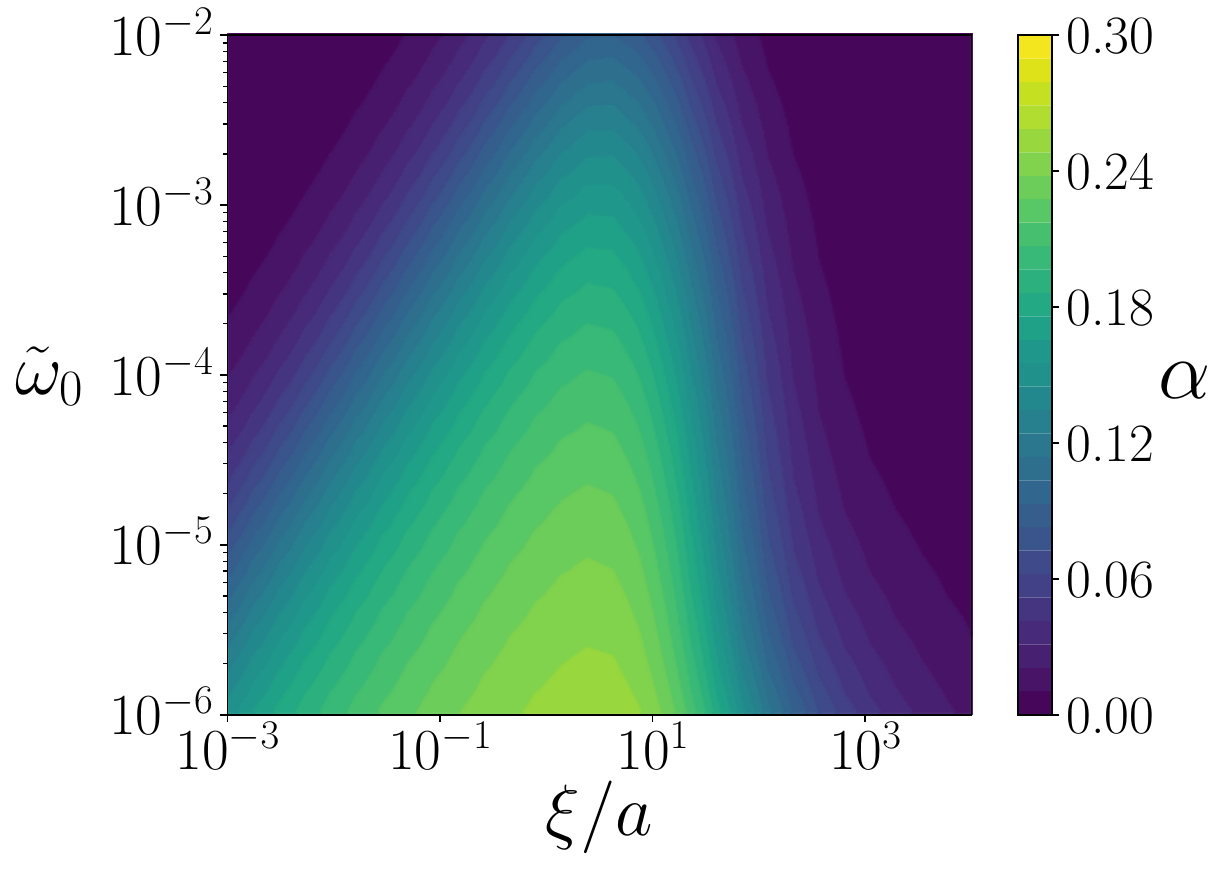}
    \caption{The logarithmic scaling coefficient $\alpha$ of the homogeneous Bogoliubov theory as a function of the relative length scale ${\xi}/{a}$, and the dimensionless gap parameter $\tilde{\omega}_0 \equiv {\omega_0 a}/{c_r}$, in the limit when $l\ll c_r/\omega_0 \ll L$.}
    \label{fig:scaling2}
\end{figure}

Since our acoustic black-hole system has to be infinite in size, it is not possible to choose a system size smaller than the length scale corresponding to the mass gap. It will therefore be necessary to consider the limit where $ l\ll c_r/\omega_0\ll L$. In this limit, the scaling remains logarithmic. However, the gap dependence will enter the prefactor $\alpha$. Fig.~\ref{fig:scaling2} shows the scaling prefactor for the two different parameters. Because of this dependence on the gap, it is more difficult to extract the universal terms for the entanglement entropy. However, it will still be possible to study the general structure, as we discuss in the following section. 

\begin{figure}[t]
    \centering
    \includegraphics[width=0.963\linewidth]{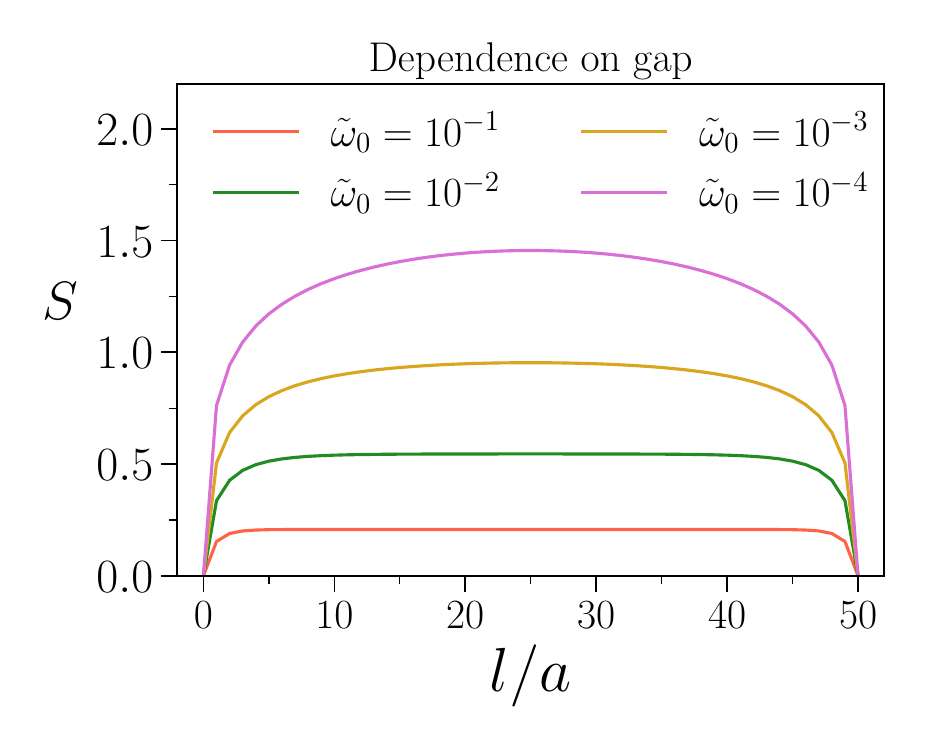}
    \caption{The entanglement entropy of a finite-size homogeneous Bogoliubov theory for different values of the dimensionless gap parameter $\tilde{\omega}_0 = {\omega_0 a}/{c_r}$ for a system of $N=50$ points, with $\xi=a$. The graph clearly shows the transition from a logarithmic to a constant behavior when the gap is increased. }
    \label{fig:mass}
\end{figure}

Finally, for subsystem sizes $l$ larger than that corresponding to the gap, $l > c_r/\omega_0$, the entanglement entropy changes its scaling behavior from logarithmic to a constant. Fig.~\ref{fig:mass} clearly shows this transition in the entanglement entropy for different values of the gap parameter. This fully agrees with the expected area law of a massive field theory with a finite correlation length and a subsystem size that is larger than that, so that in the expression for the entanglement entropy the size $l$ is cut off by the correlation length $c_r/\omega_0$. 
\newline
\newline

\subsection{Entanglement across the horizon}
\begin{figure}
    \centering
    \includegraphics[width=0.93\linewidth]{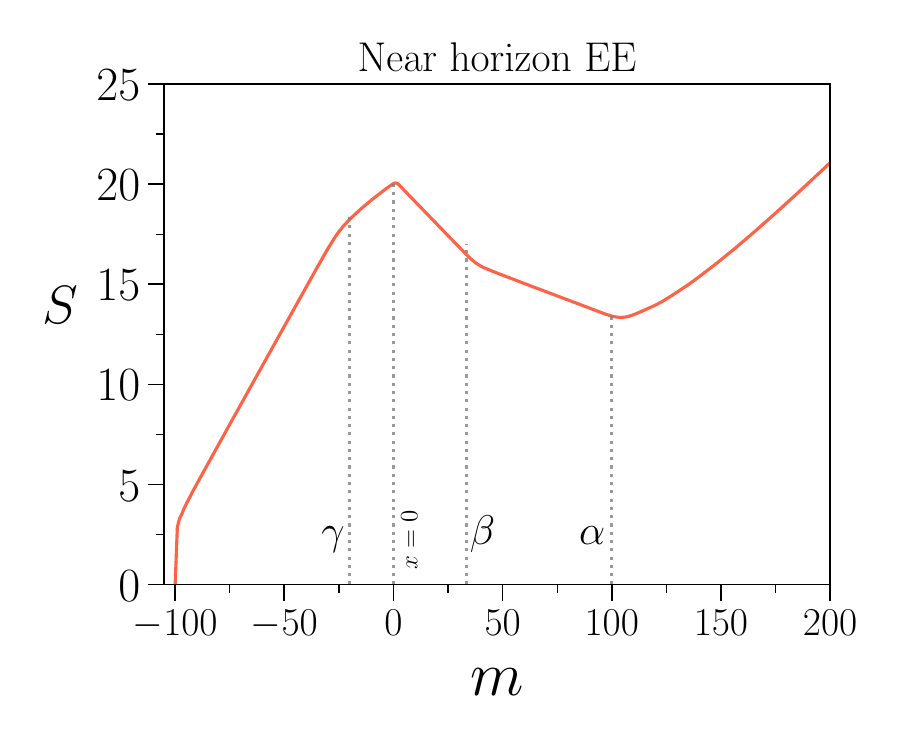}
    \caption{The entanglement entropy of the subregion $\left[-100a,ma\right]$ of the acoustic black hole as a function of $m$. The features from Fig.~\ref{fig:correlations} are denoted with dotted lines, labeled $\alpha$, $\beta$ and $\gamma$. The flow profile has ${c_l}/{|v|} = {1}/{2}$ and ${c_r}/{|v|}= {3}/{2}$, with the healing length $\xi_l = 2{a}$ and gap ${\omega_0 a}/{c_l}=5\times10^{-6}$. }
    \label{fig:doorsnee}
\end{figure}

Since the black-hole interior in our set-up contains an infinite number of points, it does not allow for the computation of the entanglement entropy of the subsystem outside the black hole, as this is a divergent quantity. It is, however, possible to study the influence of the long-range correlations on the entanglement entropy near the horizon. Fig.~\ref{fig:doorsnee} shows the entanglement entropy for a subsystem consisting of the points in $\left[-Ma,ma\right]$, with varying $m$ from $-M$ to $2M$. As can be seen, the entanglement entropy grows linearly when the boundary moves towards the horizon. At some point before reaching the horizon at $x=0$, the slope drops slightly. When the horizon is crossed, the entanglement entropy decreases. Again, after some point, the slope reduces, until it reaches a minimum at the opposite side of the black hole at $m_{min}=M$. After this, it increases again linearly. 

This behavior can be explained qualitatively using the long-distance correlations in Fig.~\ref{fig:correlations}. The initial linear growth is due to the thermal entropy of the infalling Hawking radiation as we discuss in detail in the following subsection. The first kink is accounted for by the relatively weak correlation peak $\gamma$ between the two kinds of infalling Hawking phonons. When the boundary crosses this line, these correlations start to take place within the subregion itself instead of between the subregion and its environment, leading to a smaller growth in entanglement entropy that continues until the horizon is crossed. At that point the much stronger correlations between the infalling and outgoing Hawking pairs corresponding to the correlation peaks $\alpha$ and $\beta$ start to take place within the subregion of interest. This now even leads to a drop in the entanglement entropy. Again a kink occurs when all $\beta$ correlations occur inside the subregion and only more $\alpha$ correlations can be included. The latter results in a further reduction and ultimately in a minimum in the entanglement entropy around this latter feature. From there on the entanglement entropy then increases linearly again due to the thermal entropy of the outgoing Hawking radiation. 

We thus clearly see that the subsystem sizes at which the kinks appear are determined by the ratios of the three relevant group velocities as described by Eq.~(\ref{group_vels}), and it is evident that the behavior of the entanglement entropy is highly dependent on the existence of the long-distance correlations of the Hawking phonons, and their corresponding strengths. This also suggests that these correlations are not separable and in fact include entangled degrees of freedom as expected from the tunneling nature of the pair production process \cite{Wilczek, tunneling_analogue_grav}. 

\begin{figure}
    \centering
    \includegraphics[width=0.93\linewidth]{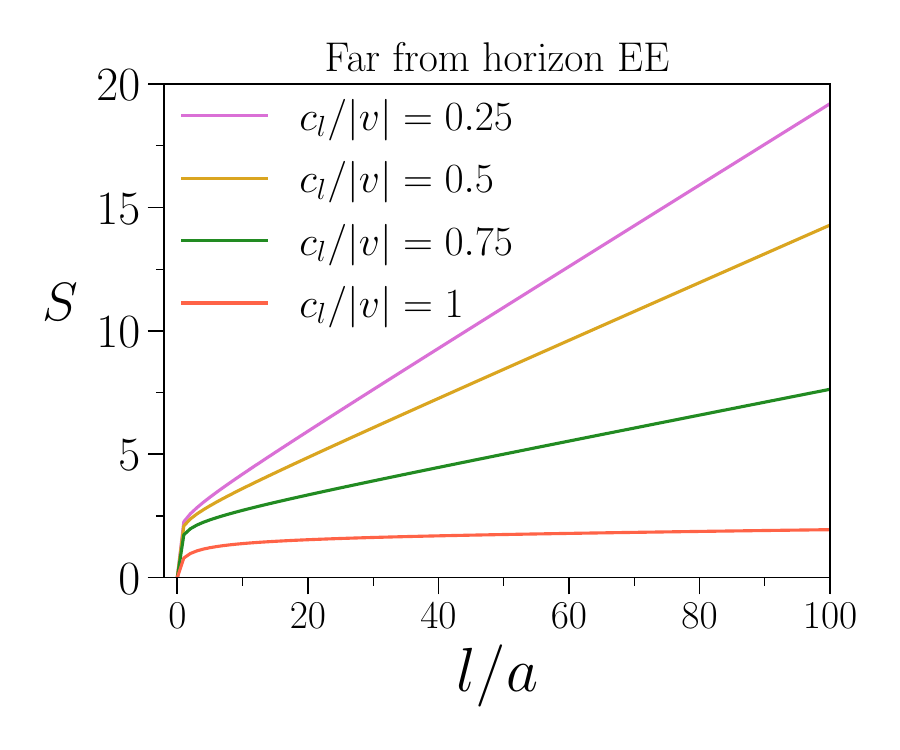}
    \caption{The entanglement entropy of the region $[10^3a,10^3a+la]$, as function of $l/a$ for different values of the speed of sound in the supersonic region $c_l$. The speed of sound in the subsonic region is given by ${c_r}/{|v|}= {3}/{2}$, the healing length $\xi_l = 2a$ and gap ${\omega_0a}/{c_l}=5\times10^{-6}$. }
    \label{fig:Farhorizon}
\end{figure}

\subsection{Volume-law scaling}
\begin{figure}
    \centering
    \includegraphics[width=0.85\linewidth]{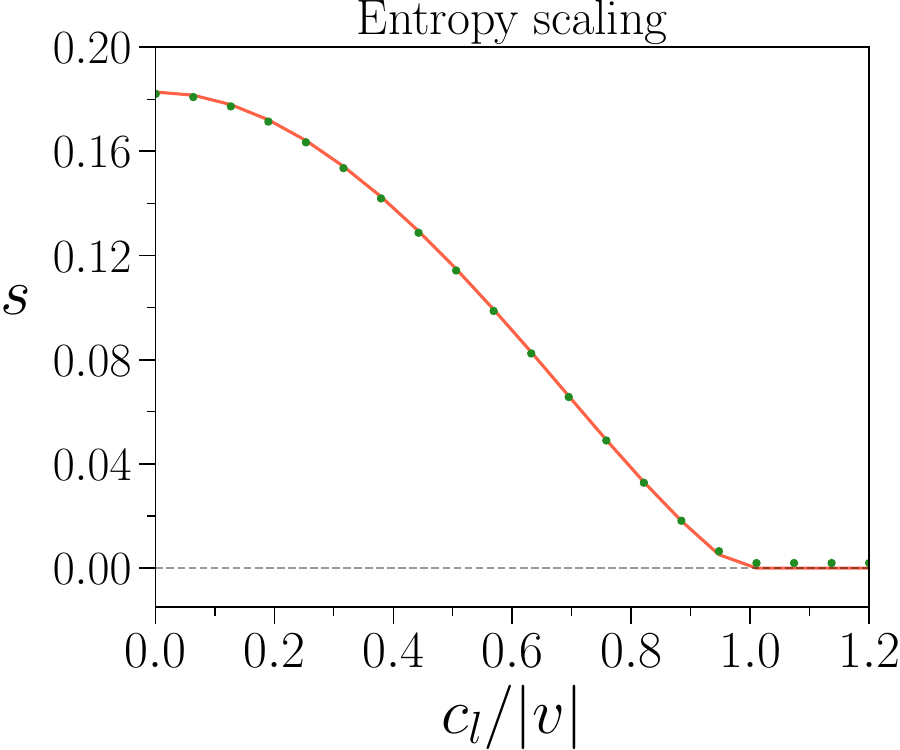}
    \caption{The entanglement entropy per lattice site $s_{th}$ in the subsonic region as function of the speed of sound in the supersonic region. The green dots denote the values computed from the correlation functions, and the red line to those computed from Eq.~(\ref{thermal entropy}). The speed of sound in the subsonic region is ${c_r}/{|v|}= {3}/{2}$, with healing length $\xi_r = 2{a}$ and gap ${\omega_0 a}/{c_r}=1/6\times10^{-5}$. At $c_l/|v| = 1$, the flow transitions from transcritical to subcritical, such that the volume law disappears. }
    \label{fig:scaling_ding}
\end{figure}
As can be seen from Fig.~\ref{fig:Farhorizon}, the entanglement entropy for subsystems far from the horizon scale linearly with subsystem size, instead of the usual logarithmic or constant scaling. Similar to the long-distance correlations, this volume-law scaling only appears for transsonic configurations, and disappears completely when the flow is everywhere subsonic, as can be concluded from Fig.~\ref{fig:scaling_ding}. Since the system is locally thermal, the scaling of the entanglement entropy can be interpreted as due to the nonzero thermal entropy density of the outgoing mode. This entropy can be computed from the scattering coefficients by considering a finite-size system with periodic boundary conditions, where the right-moving modes are occupied according to Eqs.~(\ref{occupation1}) and (\ref{occupation2}). The von Neumann entropy of such a state is computed from the occupation numbers as 
\begin{align}
S_{th} = \sum_i \left[ (1+N_i)\log(1+N_i) - N_i\log(N_i) \right],
\label{thermal entropy}
\end{align}
where $N_i = \langle a_i^\dagger a_i\rangle$ \cite{demarie2012pedagogicalintroductionentropyentanglement}. This can be computed numerically from the scattering amplitudes. The details of this computation can be found in the appendix. 

In Fig.~\ref{fig:scaling_ding} we show that the corresponding volume-law of the entanglement entropy agrees well with that of Eq.~(\ref{thermal entropy}), supporting the idea that the volume scaling is due to the local thermal effects. Subleading to the volume term, there is the usual logarithmic scaling due to the short-distance entanglement across the boundary of the subregion. Numerical computations show that the logarithmic-scaling coefficient agrees with that of the homogeneous system. This leads to the following size dependence for regions away from the horizon,
\begin{align}
S_{E} = s_{th}\frac{l}{a}+ \alpha\log\left(\frac{l}{a}\right) + c'.
\end{align}
Here $s_{th}$ is the thermal entropy per site as computed using Eq.~(\ref{thermal entropy}), and $\alpha$ is the scaling coefficients for the homogeneous Bogoliubov field in the limit when $l \ll c_l/\omega_0 \ll L$ as discussed in the previous section, which thus depends on the ratio $\xi/a$ and the gap $\omega_0$. At length scales much larger than that corresponding to the mass gap, the logarithmic scaling is replaced by a constant, but the volume-law behavior remains as 
\begin{align}
S_E = s_{th}\frac{l}{a} + c''.
\end{align}

These results suggest a general form of the size-dependent scaling of entanglement entropy for regions outside the event horizon of any black hole. The existence of a horizon introduces long-distance correlations between points inside and outside the black hole, leading to a volume-law scaling. The scaling coefficient can be computed directly from the outgoing Hawking spectrum by considering the entropy density of a field theory with thermalized right-moving modes, similar to the computation originally done by 't Hooft \cite{tHooft}. The short-distance correlations are unaffected by the black hole, such that the subleading terms in the entanglement entropy are given by that of the homogeneous theory.
\quad\\

\section{Conclusions}
In this work, the behavior of the entanglement entropy has been studied in the setting of a superfluid black-hole analogue in a Bose-Einstein condensate. By introducing a tight-binding approximation and a mass term in the equation of motion, the infrared and ultraviolet divergences are removed. This allowed the use of the correlator method for computing the entanglement entropy in the acoustic black hole. For a homogeneous system, the Bogoliubov theory shows a logarithmic scaling with subsystem size, for which the prefactor $\alpha$ transitions between the known value for conformal field theories of ${1}/{3}$ when $\xi \ll a$, and the value ${1}/{2}$ when $\xi \gg a$. The presence of a black-hole horizon introduces non-separable long-distance correlations, which lead to a volume-law scaling in the entanglement entropy far from the horizon. We have shown that this scaling is well approximated by the thermal entropy of a closed system with a spectrum of that of the black hole. Subleading to this volume term there is the usual logarithmic scaling for a massless theory due to the short-distance correlations over the boundaries.

We hope that our work may led to further developments on the entanglement entropy of analogue black holes. One interesting direction is to explicitly determine the wave function of the Hawking pairs and consider the entanglement entropy of the pairs directly. Another promising direction is the consideration of the two-dimensional radial vortex in a Bose-Einstein condensate put forward by Liao {\it et al.} \cite{PhysRevA.99.023850}, and which in the hydrodynamic regime recovers a true singular and rotationally symmetric Schwarzschild black hole. Finally, anti-magnon spintronic devices \cite{artim} are of interest as they have no hydrodynamic regime with a metric and the individual homogeneous regions also have no entanglement entropy to start with in the fully polarized ferromagnetic state. 

In the final stages of writing up our work we became aware of the preprint by Chandran and Fisher \cite{CF}, who do not determine the entanglement entropy of an acoustic black hole in a Bose-Einstein condensate, but also find a volume law for its entanglement negativity in the conformal field-theory limit using completely different methods.  

\section{Acknowledgments}
We thank the master students Hendrik Snijder, Ken de Ruiter, Mike Hakvoort, and Pablo Vis for preliminary work on this topic, and Remco Bus for helpful discussions. We are very grateful to Stefan Vandoren and Rembert Duine for various useful discussions on black holes in quantum gravity and on anti-magnonics, respectively. 

This work is supported by the Delta-ITP consortium, which is part of the Netherlands Organisation for Scientific Research (NWO). We also acknowledge the research program “Materials for the Quantum Age” (QuMat) for financial support. This program (registration number 024.005.006) is part of the Gravitation program financed by the Dutch Ministry of Education, Culture and Science (OCW). 

\section*{Appendix}
The matching conditions in the different regimes are discussed here in the appendix. The amplitudes are given by the scattering matrix $S$ for propagating modes, and will be labeled by $A$ and $B$ for the evanescent modes in the ingoing and outgoing basis respectively. The different propagating and evanescent momenta in the sub- and supersonic regions are shown in table \ref{table}. The entries in the vector notation are given by the momenta $k_{i,-}, k_{i,+}, k_{o,-}, k_{o,+}$ respectively from top to bottom. 
\begin{table}[b]
\begin{tabular}{|c|c|c|c|c|c|}
        \hline
        $k$ & I & II & III & IV\\
        \hline
        $k^l_{i,-}$ & $\checkmark$  & $\checkmark$ & $\times$ & $\times$\\
        $k^l_{i,+}$ & $\checkmark$  & $\checkmark$ & $\checkmark$ & $\times$\\
        $k^l_{o,-}$ & $\checkmark$  & $\checkmark$ & ev & ev\\
        $k^l_{o,+}$ & $\checkmark$  & $\checkmark$ & $\checkmark$ & ev\\
        \hline
        $k^r_{i,-}$ & $\times$  & $\times$ & $\times$ & $\times$\\
        $k^r_{i,+}$ & $\times$  & $\checkmark$ & $\checkmark$ & $\checkmark$\\
        $k^r_{o,-}$ & ev & ev & ev & ev\\
        $k^r_{o,+}$ & ev & $\checkmark$ & $\checkmark$ & $\checkmark$\\
        \hline
\end{tabular}
\caption{A table that labels which modes are propagating in which frequency regime. A checkmark denotes a propagating mode, ev a decaying evanescent mode, and a cross a growing evanescent mode, which is not allowed. }
\label{table}
\end{table}
\subsection{Regime I}
In the first region below the mass gap $\omega < \omega_{\text{gap}}$, there are only propagating modes in the supersonic region, and evanescent modes in the subsonic region. The solutions thus represent full reflections off the horizon. The incoming momenta are given by $k_{i,-}^l$ and $k_{i,+}^l$, and the outgoing ones by $k_{o,+}^l$, $k_{o,-}^l$. The evanescent modes in the subsonic region are given by the complex continuation of $k^r_{o,+}$ and the complex mode $k^r_{o,-}$, and do not contribute to the scattering matrix. This leads to the following matching conditions for the {\it in} and {\it out} basis modes,
\vspace{-25pt}
\subsubsection{in-basis}
\vspace{-25pt}
\begin{align}
\begin{pmatrix}
1\\
0\\
S_{l,-}^{l,-}\\
S_{l,+}^{l,-}
\end{pmatrix}= M \begin{pmatrix}
0\\
0\\
A_{r,-}^{l,-}\\
A_{r,+}^{l,-}
\end{pmatrix},
\quad \begin{pmatrix}
0\\
1\\
S_{l,-}^{l,+}\\
S_{l,+}^{l,+}
\end{pmatrix}= M \begin{pmatrix}
0\\
0\\
A_{r,-}^{l,+}\\
A_{r,+}^{l,+}\\
\end{pmatrix}.\nonumber
\end{align}

\vspace{-20pt}
\subsubsection{out-basis}
\vspace{-20pt}

\begin{align}
\begin{pmatrix}
{S^{l,-}_{l,-}}^*\\
{S^{l,+}_{l,-}}^*\\
1\\
0
\end{pmatrix}= M \begin{pmatrix}
0\\
0\\
{B_{l,-}^{r,-}}\\
{B_{l,-}^{r,+}}
\end{pmatrix},\quad \begin{pmatrix}
{S^{l,-}_{l,+}}^*\\
{S^{l,+}_{l,+}}^*\\
0\\
1
\end{pmatrix}= M \begin{pmatrix}
0\\
0\\
{B_{l,+}^{r,-}}\\
{B_{l,+}^{r,+}}
\end{pmatrix}. \nonumber
\end{align}

\subsection{Regime II}

In the second regime there are four real momenta in the supersonic region, and two in the subsonic region. There are three different possible incoming modes, with momenta $k^l_{i,-}$, $k^l_{i,+}$ and $k_{i,+}^r$, and three outgoing modes with momenta $k^l_{o,-}$, $k^l_{o,+}$, $k^r_{o,+}$. Only the complex evanescent mode $k^r_{o,-}$ exists. The matching equations are now given by

\vspace{-20pt}
\subsubsection{in-basis}
\vspace{-20pt}

\begin{align}
&
\begin{pmatrix}
1\\
0\\
S_{l,-}^{l,-}\\
S_{l,+}^{l,-}
\end{pmatrix}= M \begin{pmatrix}
0\\
0\\
A_{r,-}^{l,-}\\
S^{l,-}_{r,+}\\
\end{pmatrix},
\quad \begin{pmatrix}
0\\
1\\
S_{l,-}^{l,+}\\
S_{l,+}^{l,+}
\end{pmatrix}= M \begin{pmatrix}
0\\
0\\
A_{r,-}^{l,+}\\
S_{r,+}^{l,+}
\end{pmatrix},\nonumber
\\& \begin{pmatrix}
0\\
0\\
S_{l,-}^{r,+}\\
S_{l,+}^{r,+}
\end{pmatrix}= M \begin{pmatrix}
0\\
1\\
A_{r,-}^{r,+}\\
S_{r,+}^{r,+}
\end{pmatrix}.\nonumber
\end{align}

\vspace{-20pt}
\subsubsection{out-basis}
\vspace{-20pt}

\begin{align}
& \begin{pmatrix}
{S^{l,-}_{l,-}}^*\\
{S^{l,+}_{l,-}}^*\\
1\\
0
\end{pmatrix}= M \begin{pmatrix}
0\\
{S^{r,+}_{l,-}}^*\\
B_{l,-}^{r,-}\\
0
\end{pmatrix},
 \quad \begin{pmatrix}
{S^{l,-}_{l,+}}^*\\
{S^{l,+}_{l,+}}^*\\
0\\
1
\end{pmatrix}= M \begin{pmatrix}
0\\
{S^{r,+}_{l,+}}^*\\
B_{l,+}^{r,-}\\
0
\end{pmatrix},
\nonumber \\ & \begin{pmatrix}
{S^{l,-}_{r,+}}^*\\
{S^{l,+}_{r,+}}^*\\
0\\
0
\end{pmatrix}= M \begin{pmatrix}
0\\
{S^{r,+}_{r,+}}^*\\
B_{r,+}^{r,-}\\
1
\end{pmatrix}.\nonumber
\end{align}

\subsection{Regime III}
Regime III has two propagating modes in the supersonic and subsonic regions. There are two different solutions in the ingoing basis with momentum $k_{i,+}^l$ and $k^r_{i,+}$, and two in the outgoing with $k_{o,+}^l$ and $k^r_{o,+}$. On both the subsonic and supersonic regions there is one evanescent mode, namely $k^{l}_{o,-}$ and $k^r_{o,-}$.

\vspace{-20pt}
\subsubsection{in-basis}
\vspace{-20pt}

\begin{align}
\begin{pmatrix}
0\\
1\\
A^{l,+}_{l,-}\\
S^{l,+}_{l,+}
\end{pmatrix}= M \begin{pmatrix}
0\\
0\\
A_{r,-}^{l,+}\\
S_{r,+}^{l,+}
\end{pmatrix},
\quad \begin{pmatrix}
0\\
0\\
A_{l,-}^{r,+}\\
S_{l,+}^{r,+}
\end{pmatrix}= M \begin{pmatrix}
0\\
1\\
A_{r,-}^{r,+}\\
S^{r,+}_{r,+}
\end{pmatrix}.\nonumber
\end{align}

\vspace{-20pt}
\subsubsection{out-basis}
\vspace{-20pt}

\begin{align}
\begin{pmatrix}
0\\
{S_{l,+}^{l,+}}^*\\
B_{l,+}^{l,-}\\
1
\end{pmatrix}= M \begin{pmatrix}
0\\
{S_{l,+}^{r,+}}^*\\
B_{l,+}^{r,-}\\
0
\end{pmatrix},
\quad \begin{pmatrix}
0\\
{S_{r,+}^{l,+}}^*\\
B_{r,+}^{l,-}\\
0
\end{pmatrix}= M \begin{pmatrix}
0\\
{S^{r,+}_{r,+}}^*\\
B^{r,-}_{r,+}\\
1
\end{pmatrix}.\nonumber
\end{align}

\subsection{Regime IV}
Finally, in the last regime there are only two real solutions left in the subsonic region. This leads to only one ingoing mode with momentum $k_{i,+}^r$ and one outgoing mode with $k^r_{o,+}$. The decaying evanescent modes are given by $k_{o,+}^l$, $k_{o,-}^l$ and $k_{o,-}^r$.

\vspace{-15pt}
\subsubsection{in-basis}
\vspace{-10pt}

\begin{align}
\begin{pmatrix}
0\\
0\\
A_{l,-}^{r,+}\\
A_{l,+}^{r,+}
\end{pmatrix}= M \begin{pmatrix}
0\\
1\\
A_{r,-}^{r,+}\\
S_{r,+}^{r,+}
\end{pmatrix}\nonumber.
\end{align}

\vspace{-15pt}
\subsubsection{out-basis}
\vspace{-10pt}

\begin{align}
\begin{pmatrix}
0\\
0\\
B_{r,+}^{l,-}\\
B_{r,+}^{l,+}
\end{pmatrix}= M \begin{pmatrix}
0\\
{S_{r,+}^{r,+}}^*\\
B_{r,+}^{r,-}\\
1
\end{pmatrix}\nonumber
\end{align}

\subsection{Entropy computation}
As discussed, the thermal entropy density can be computed by considering a finite-size system with periodic boundary conditions, for which the right moving modes are occupied according to Eq.~(\ref{occupation2}). We consider only the subsonic region. However, a similar computation can be done for the supersonic region. For a homogeneous subsonic system of $N$ lattice points, there are $N$ sets of mode functions $u_k (x,t) = U_k\exp\left({2\pi ik x}/{Na} - i \omega_k t\right)$, with $k=0,..,N-1$. The frequency $\omega_k$ is determined from the dispersion relation in Eq.~(\ref{disp}). The right-moving modes are given by those with positive group velocity ${d\omega}/{dk} >0$. By occupying these modes with the computed amplitudes from $N_{r,+}$ in Eq.~(\ref{occupation2}), the von Neumann entropy can be computed to be equal to 
\begin{align}
S = \sum_{k,v_g>0} \left[(1+N_k)\log(1+N_k) - N_k\log(N_k)\right],
\end{align}
where the sum is over all modes with positive group velocity and $N_k = |S^{l,-}_{r,+}(\omega_k)|^2$. By dividing by the total number if sites $N$ in the system, we obtain the entropy per site $s = S/N$. Preferably, a large number of lattice points $N \simeq 10^3$ is chosen to increase accuracy. 

\subsection{Correlators}
The momentum-momentum and momentum-field correlations can be computed similarly to Eq.~(\ref{corr_real}). For the $\pi -\pi$ correlations, we obtain
\begin{align}
&4\langle \pi(x)\pi(y)\rangle_{in}  =  \sum_{i\in\{l+,r\}} \int_0^{\omega^r_{m}} d\omega \left[u_{\omega}^i(x) - v_{\omega}^i(x)\right] \nonumber\\
& \times \left[{u_{\omega}^i}^*(y) - {v_{\omega}^i}^*(y)\right] + \int_0^{\omega^r_{m}} d\omega \left[{v_{\omega}^{l,-}}^*(x) - {u_{\omega}^{l,-}}^*(x)\right] \nonumber\\
& \times \left[v_{\omega}^{l,-}(y) - u_{\omega}^{l,-}(y)\right], 
\end{align}
and for the momentum-field correlator
\begin{align}
&2i\langle \pi(x)\rho(y)\rangle_{in}  = \sum_{i\in\{l+,r\}} \int_0^{\omega^r_{m}} d\omega \left[u_{\omega}^i(x) - v_{\omega}^i(x)\right] \nonumber\\
& \times \left[{u_{\omega}^i}^*(y) + {v_{\omega}^i}^*(y)\right] + \int_0^{\omega^r_{m}} d\omega \left[{v_{\omega}^{l,-}}^*(x) - {u_{\omega}^{l,-}}^*(x)\right] \nonumber\\
& \times \left[v_{\omega}^{l,-}(y) + u_{\omega}^{l,-}(y)\right] = 2i\langle \rho(y)\pi(x)\rangle_{in}^*.
\end{align}
Note that these definitions imply that $\left[ \rho(x),\pi(y)\right] = 2\text{Im}\left[\langle \rho(x)\pi(y)\rangle\right]$. A subsequent computation, using the completeness relations based on Eq.~(\ref{norm_aangepast}), then shows that this leads to the correct commutation relations, i.e., $\left[ \rho(x),\pi(y)\right] = i\delta(x-y)$.

\subsection{Oscillations in supersonic region and pair entanglement}
\begin{figure}[b]
    \centering
    \includegraphics[width=1\linewidth]{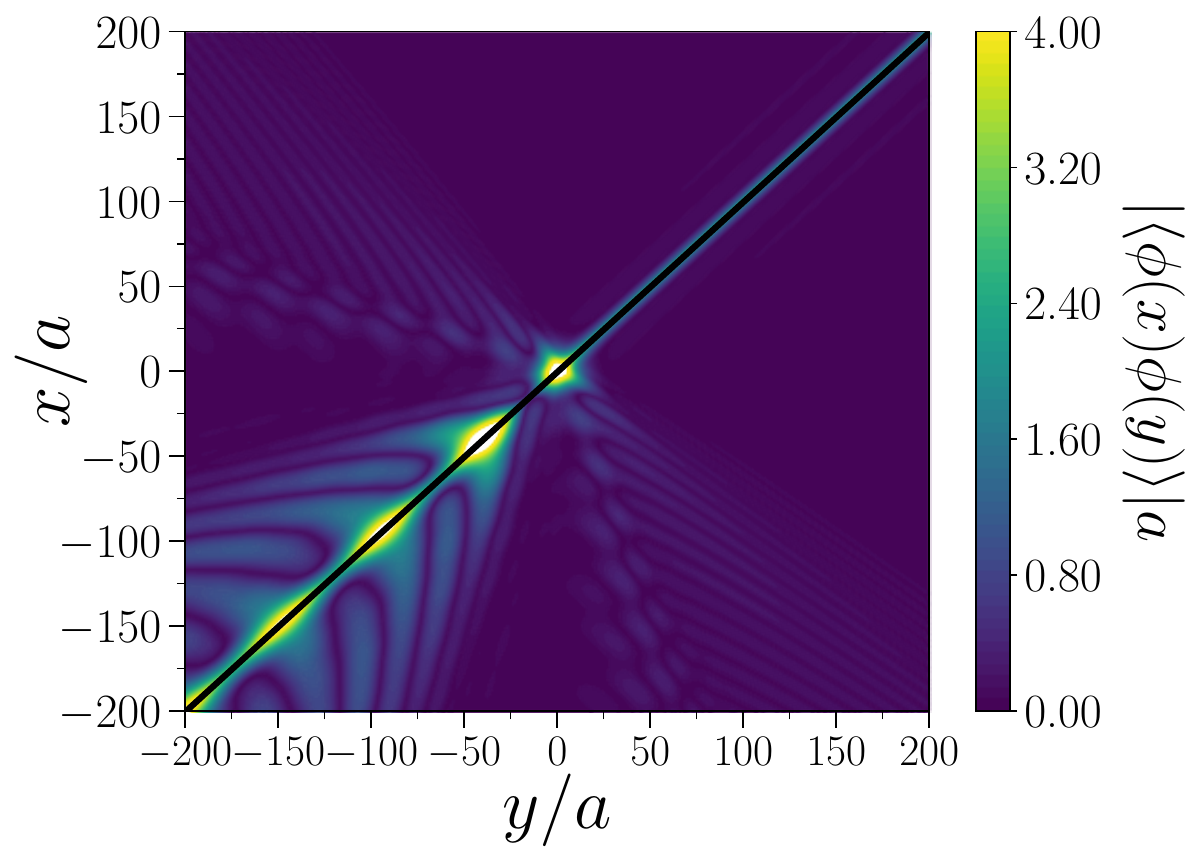}
    \caption{The $\phi - \phi$ correlation function of the acoustic black hole, with a gap $\omega_0 a /c_l = 5\times10^{-3}$. The flow profile has $c_l/|v| = 1/2$ and $c_r/|v| = 3/2$, with healing length $\xi_l=a$. }
    \label{fig:oscillations}
\end{figure}

While the main features in the $\rho - \rho$ correlator do not depend very strongly on the gap parameter, it should be noted that the anomalous correlator of the original field, i.e., $\langle \phi(x) \phi(y)\rangle$, does however. In Fig.~\ref{fig:oscillations} we show the absolute value of these correlations near the horizon. The dependence is mostly visible in the supersonic region. Due to the high occupation of low-energy modes here, the correlations are mostly determined by the full reflections in regime I. The wavelength of these oscillations is determined by the interference between the modes $k^l_{o,+}$ and $k^l_{o,-}$, such that $\lambda = {2\pi}/({k^l_{o,+} - k^l_{o,-}})$.

Note that the discussion in this work has not explicitly considered the entanglement entropy between the individual phonon pairs, as this would require a different approach of computation and the determination of the pair wave function. We suspect it might be possible to interpret the anomalous $\phi-\phi$ correlator as a two-particle wave function, after which the entanglement between Hawking pairs might be computed and compared with the full entanglement entropy determined here. This is, however, beyond the scope of the present paper. 

\bibliography{refs}

@article{Recati_2009,
   title={Bogoliubov theory of acoustic Hawking radiation in Bose-Einstein condensates},
   volume={80},
   ISSN={1094-1622},
   url={http://dx.doi.org/10.1103/PhysRevA.80.043603},
   DOI={10.1103/physreva.80.043603},
   number={4},
   journal={Physical Review A},
   publisher={American Physical Society (APS)},
   author={Recati, A. and Pavloff, N. and Carusotto, I.},
   year={2009},
   month=oct }

@article{Calabrese_2009,
   title={Entanglement entropy and conformal field theory},
   volume={42},
   ISSN={1751-8121},
   url={http://dx.doi.org/10.1088/1751-8113/42/50/504005},
   DOI={10.1088/1751-8113/42/50/504005},
   number={50},
   journal={Journal of Physics A: Mathematical and Theoretical},
   publisher={IOP Publishing},
   author={Calabrese, Pasquale and Cardy, John},
   year={2009},
   month=dec, pages={504005} }

@article{Bekenstein,
  title = {Black Holes and Entropy},
  author = {Bekenstein, Jacob D.},
  journal = {Phys. Rev. D},
  volume = {7},
  issue = {8},
  pages = {2333--2346},
  numpages = {0},
  year = {1973},
  month = {Apr},
  publisher = {American Physical Society},
  doi = {10.1103/PhysRevD.7.2333},
  url = {https://link.aps.org/doi/10.1103/PhysRevD.7.2333}
}

@article{Hawking,
author = {S. W. Hawking},
title = {{Particle creation by black holes}},
volume = {43},
journal = {Communications in Mathematical Physics},
number = {3},
publisher = {Springer},
pages = {199 -- 220},
year = {1975},
}

@article{Bombelli,
  title = {Quantum source of entropy for black holes},
  author = {Bombelli, Luca and Koul, Rabinder K. and Lee, Joohan and Sorkin, Rafael D.},
  journal = {Phys. Rev. D},
  volume = {34},
  issue = {2},
  pages = {373--383},
  numpages = {0},
  year = {1986},
  month = {Jul},
  publisher = {American Physical Society},
  doi = {10.1103/PhysRevD.34.373},
  url = {https://link.aps.org/doi/10.1103/PhysRevD.34.373}
}

@article{tHooft,
    author = "'t Hooft, Gerard",
    title = "{On the Quantum Structure of a Black Hole}",
    reportNumber = "Print-84-0924 (UTRECHT)",
    doi = "10.1016/0550-3213(85)90418-3",
    journal = "Nucl. Phys. B",
    volume = "256",
    pages = "727--745",
    year = "1985"
}

@article{Steinhauer,
   title={Observation of self-amplifying Hawking radiation in an analogue black-hole laser},
   volume={10},
   ISSN={1745-2481},
   url={http://dx.doi.org/10.1038/NPHYS3104},
   DOI={10.1038/nphys3104},
   number={11},
   journal={Nature Physics},
   publisher={Springer Science and Business Media LLC},
   author={Steinhauer, Jeff},
   year={2014},
   month=Oct, pages={864–869} }

@article{Fischer,
   title={Existence of steady-state black hole analogs in finite quasi-one-dimensional Bose-Einstein condensates},
   volume={105},
   ISSN={2470-0029},
   url={http://dx.doi.org/10.1103/PhysRevD.105.124066},
   DOI={10.1103/physrevd.105.124066},
   number={12},
   journal={Physical Review D},
   publisher={American Physical Society (APS)},
   author={Ribeiro, Caio C. Holanda and Baak, Sang-Shin and Fischer, Uwe R.},
   year={2022},
   month=June }

@article{Unruh,
  title = {Experimental Black-Hole Evaporation?},
  author = {Unruh, W. G.},
  journal = {Phys. Rev. Lett.},
  volume = {46},
  issue = {21},
  pages = {1351--1353},
  numpages = {0},
  year = {1981},
  month = {May},
  publisher = {American Physical Society},
  doi = {10.1103/PhysRevLett.46.1351},
  url = {https://link.aps.org/doi/10.1103/PhysRevLett.46.1351}
}

@article{srednicki,
   title={Entropy and area},
   volume={71},
   ISSN={0031-9007},
   url={http://dx.doi.org/10.1103/PhysRevLett.71.666},
   DOI={10.1103/physrevlett.71.666},
   number={5},
   journal={Physical Review Letters},
   publisher={American Physical Society (APS)},
   author={Srednicki, Mark},
   year={1993},
   month=Aug, pages={666–669} }

@article{symm_breaking,
   title={Degeneracies and symmetry breaking in pseudo-Hermitian matrices},
   volume={5},
   ISSN={2643-1564},
   url={http://dx.doi.org/10.1103/PhysRevResearch.5.023035},
   DOI={10.1103/physrevresearch.5.023035},
   number={2},
   journal={Physical Review Research},
   publisher={American Physical Society (APS)},
   author={Melkani, Abhijeet},
   year={2023},
   month=Apr }

@book{Henk_boek,
    author = "Stoof, Henk T. C. and Gubbels, Koos B. and Dickerscheid, Dennis B. M.",
    title = "{Ultracold Quantum Fields}",
    doi = "10.1007/978-1-4020-8763-9",
    isbn = "978-1-4020-8762-2, 978-1-4020-8763-9",
    publisher = "Springer",
    address = "Berlin, Germany",
    series = "Theoretical and Mathematical Physics",
    year = "2009"
}

@article{Jeff2,
   title={Observation of quantum Hawking radiation and its entanglement in an analogue black hole},
   volume={12},
   ISSN={1745-2481},
   url={http://dx.doi.org/10.1038/nphys3863},
   DOI={10.1038/nphys3863},
   number={10},
   journal={Nature Physics},
   publisher={Springer Science and Business Media LLC},
   author={Steinhauer, Jeff},
   year={2016},
   month=Aug, pages={959–965} }

@misc{demarie2012pedagogicalintroductionentropyentanglement,
      author={Tommaso F. Demarie},
      year={2012},
      eprint={1209.2748},
      archivePrefix={arXiv},
      primaryClass={quant-ph},
      url={https://arxiv.org/abs/1209.2748}, 
}

@article{EE1,
  title = {Horizon entanglement area law from regular black hole thermodynamics},
  author = {Belfiglio, Alessio and Chandran, S. Mahesh and Luongo, Orlando and Mancini, Stefano},
  journal = {Phys. Rev. D},
  volume = {111},
  issue = {2},
  pages = {024013},
  numpages = {16},
  year = {2025},
  month = {Jan},
  publisher = {American Physical Society},
  doi = {10.1103/PhysRevD.111.024013},
  url = {https://link.aps.org/doi/10.1103/PhysRevD.111.024013}
}

@article{EE2,
    author = "Huang, Shifeng and Fang, Xiongjun and Jing, Jiliang",
    title = "{Numerical calculation of the entanglement entropy for...}",
    doi = "10.1007/s10714-018-2394-0",
    journal = "Gen. Rel. Grav.",
    volume = "50",
    number = "6",
    pages = "70",
    year = "2018"
}

@article{EE3,
   title={Quantum entanglement and Hawking temperature},
   volume={76},
   ISSN={1434-6052},
   url={http://dx.doi.org/10.1140/epjc/s10052-016-4241-3},
   DOI={10.1140/epjc/s10052-016-4241-3},
   number={7},
   journal={The European Physical Journal C},
   publisher={Springer Science and Business Media LLC},
   author={Kumar, S. Santhosh and Shankaranarayanan, S.},
   year={2016},
   month=July }

@article{Symplectic,
   title={Symplectic invariants, entropic measures and correlations of Gaussian states},
   volume={37},
   ISSN={1361-6455},
   url={http://dx.doi.org/10.1088/0953-4075/37/2/L02},
   DOI={10.1088/0953-4075/37/2/l02},
   number={2},
   journal={Journal of Physics B: Atomic, Molecular and Optical Physics},
   publisher={IOP Publishing},
   author={Serafini, Alessio and Illuminati, Fabrizio and Siena, Silvio De},
   year={2003},
   month=Dec, pages={L21–L28} }

@article{Water,
   title={Measurement of Stimulated Hawking Emission in an Analogue System},
   volume={106},
   ISSN={1079-7114},
   url={http://dx.doi.org/10.1103/PhysRevLett.106.021302},
   DOI={10.1103/physrevlett.106.021302},
   number={2},
   journal={Physical Review Letters},
   publisher={American Physical Society (APS)},
   author={Weinfurtner, Silke and Tedford, Edmund W. and Penrice, Matthew C. J. and Unruh, William G. and Lawrence, Gregory A.},
   year={2011},
   month=Jan }

@article{artim,
  title = {Entangled magnon-pair generation in a driven synthetic antiferromagnet},
  author = {Bassant, A. L. and Regout, M. E. Y. and Harms, J. S. and Duine, R. A.},
  journal = {Phys. Rev. B},
  volume = {110},
  issue = {9},
  pages = {094441},
  numpages = {12},
  year = {2024},
  month = {Sep},
  publisher = {American Physical Society},
  doi = {10.1103/PhysRevB.110.094441},
  url = {https://link.aps.org/doi/10.1103/PhysRevB.110.094441}
}

@article{Casini&Huerta,
   title={Entanglement entropy in free quantum field theory},
   volume={42},
   ISSN={1751-8121},
   url={http://dx.doi.org/10.1088/1751-8113/42/50/504007},
   DOI={10.1088/1751-8113/42/50/504007},
   number={50},
   journal={Journal of Physics A: Mathematical and Theoretical},
   publisher={IOP Publishing},
   author={Casini, H and Huerta, M},
   year={2009},
   month=Dec, pages={504007} }

@article{Wilczek,
   title={Hawking Radiation As Tunneling},
   volume={85},
   ISSN={1079-7114},
   url={http://dx.doi.org/10.1103/PhysRevLett.85.5042},
   DOI={10.1103/physrevlett.85.5042},
   number={24},
   journal={Physical Review Letters},
   publisher={American Physical Society (APS)},
   author={Parikh, Maulik K. and Wilczek, Frank},
   year={2000},
   month=Dec, pages={5042–5045} }

@article{PhysRevA.99.023850,
  title = {Proposal for an analog Schwarzschild black hole in condensates of light},
  author = {Liao, L. and van der Wurff, E. C. I. and van Oosten, D. and Stoof, H. T. C.},
  journal = {Phys. Rev. A},
  volume = {99},
  issue = {2},
  pages = {023850},
  numpages = {6},
  year = {2019},
  month = {Feb},
  publisher = {American Physical Society},
  doi = {10.1103/PhysRevA.99.023850},
  url = {https://link.aps.org/doi/10.1103/PhysRevA.99.023850}
}

@misc{CF,
      author={S. Mahesh Chandran and Uwe R. Fischer},
      year={2026},
      eprint={2604.02075},
      archivePrefix={arXiv},
      primaryClass={gr-qc},
      url={https://arxiv.org/abs/2604.02075}, 
}

@article{Strominger_1996,
   title={Microscopic origin of the Bekenstein-Hawking entropy},
   volume={379},
   ISSN={0370-2693},
   url={http://dx.doi.org/10.1016/0370-2693(96)00345-0},
   DOI={10.1016/0370-2693(96)00345-0},
   number={1-4},
   journal={Physics Letters B},
   publisher={Elsevier BV},
   author={Strominger, Andrew and Vafa, Cumrun},
   year={1996},
   month=June, pages={99–104} }

@article{Susskind:2006oza,
    author = "Susskind, Leonard",
    title = "{The paradox of quantum black holes}",
    doi = "10.1038/nphys429",
    journal = "Nature Phys.",
    volume = "2",
    number = "10",
    pages = "665--677",
    year = "2006"
}

@article{Susskind_holo,
   title={The world as a hologram},
   volume={36},
   ISSN={1089-7658},
   url={http://dx.doi.org/10.1063/1.531249},
   DOI={10.1063/1.531249},
   number={11},
   journal={Journal of Mathematical Physics},
   publisher={AIP Publishing},
   author={Susskind, Leonard},
   year={1995},
   month=Nov, pages={6377–6396} }

@misc{hooft2009dimensionalreductionquantumgravity, 
      author={G. 't Hooft},
      year={2009},
      eprint={gr-qc/9310026},
      archivePrefix={arXiv},
      primaryClass={gr-qc},
      url={https://arxiv.org/abs/gr-qc/9310026}, 
}

@article{Jacobson_1998,
   title={Event horizons and ergoregions in<mml:math xmlns:mml=“http://www.w3.org/1998/Math/MathML” display=“inline”><mml:mrow><mml:msup><mml:mrow/><mml:mrow><mml:mn>3</mml:mn></mml:mrow></mml:msup></mml:mrow><mml:mi mathvariant=“normal”>He</mml:mi></mml:math>},
   volume={58},
   ISSN={1089-4918},
   url={http://dx.doi.org/10.1103/PhysRevD.58.064021},
   DOI={10.1103/physrevd.58.064021},
   number={6},
   journal={Physical Review D},
   publisher={American Physical Society (APS)},
   author={Jacobson, T. A. and Volovik, G. E.},
   year={1998},
   month=Aug }

@article{Solnyshkov_2011,
   title={Black holes and wormholes in spinor polariton condensates},
   volume={84},
   ISSN={1550-235X},
   url={http://dx.doi.org/10.1103/PhysRevB.84.233405},
   DOI={10.1103/physrevb.84.233405},
   number={23},
   journal={Physical Review B},
   publisher={American Physical Society (APS)},
   author={Solnyshkov, D. D. and Flayac, H. and Malpuech, G.},
   year={2011},
   month=Dec }

@article{Horstmann_2010,
   title={Hawking Radiation from an Acoustic Black Hole on an Ion Ring},
   volume={104},
   ISSN={1079-7114},
   url={http://dx.doi.org/10.1103/PhysRevLett.104.250403},
   DOI={10.1103/physrevlett.104.250403},
   number={25},
   journal={Physical Review Letters},
   publisher={American Physical Society (APS)},
   author={Horstmann, B. and Reznik, B. and Fagnocchi, S. and Cirac, J. I.},
   year={2010},
   month=June }

@article{light,
  title = {All-optical event horizon in an optical analog of a Laval nozzle},
  author = {Elazar, M. and Fleurov, V. and Bar-Ad, S.},
  journal = {Phys. Rev. A},
  volume = {86},
  issue = {6},
  pages = {063821},
  numpages = {5},
  year = {2012},
  month = {Dec},
  publisher = {American Physical Society},
  doi = {10.1103/PhysRevA.86.063821},
  url = {https://link.aps.org/doi/10.1103/PhysRevA.86.063821}
}

@article{Volovik_weyl_fermions,
   title={Black hole and hawking radiation by type-II Weyl fermions},
   volume={104},
   ISSN={1090-6487},
   url={http://dx.doi.org/10.1134/S0021364016210050},
   DOI={10.1134/s0021364016210050},
   number={9},
   journal={JETP Letters},
   publisher={Pleiades Publishing Ltd},
   author={Volovik, G. E.},
   year={2016},
   month=Nov, pages={645–648} }

@article{Giovanazzi_2005,
   title={Hawking Radiation in Sonic Black Holes},
   volume={94},
   ISSN={1079-7114},
   url={http://dx.doi.org/10.1103/PhysRevLett.94.061302},
   DOI={10.1103/physrevlett.94.061302},
   number={6},
   journal={Physical Review Letters},
   publisher={American Physical Society (APS)},
   author={Giovanazzi, S.},
   year={2005},
   month=Feb }

@article{light_dispersive_media,
  title = {Relativistic Effects of Light in Moving Media with Extremely Low Group Velocity},
  author = {Leonhardt, U. and Piwnicki, P.},
  journal = {Phys. Rev. Lett.},
  volume = {84},
  issue = {5},
  pages = {822--825},
  numpages = {0},
  year = {2000},
  month = {Jan},
  publisher = {American Physical Society},
  doi = {10.1103/PhysRevLett.84.822},
  url = {https://link.aps.org/doi/10.1103/PhysRevLett.84.822}
}

@article{Magnons,
   title={Magnonic Black Holes},
   volume={118},
   ISSN={1079-7114},
   url={http://dx.doi.org/10.1103/PhysRevLett.118.061301},
   DOI={10.1103/physrevlett.118.061301},
   number={6},
   journal={Physical Review Letters},
   publisher={American Physical Society (APS)},
   author={Roldán-Molina, A. and Nunez, Alvaro S. and Duine, R. A.},
   year={2017},
   month=Feb }

@article{Ryu_2006,
   title={Holographic Derivation of Entanglement Entropy from the anti–de Sitter Space/Conformal Field Theory Correspondence},
   volume={96},
   ISSN={1079-7114},
   url={http://dx.doi.org/10.1103/PhysRevLett.96.181602},
   DOI={10.1103/physrevlett.96.181602},
   number={18},
   journal={Physical Review Letters},
   publisher={American Physical Society (APS)},
   author={Ryu, Shinsei and Takayanagi, Tadashi},
   year={2006},
   month=May }

@article{tunneling_analogue_grav,
   title={Tunneling method for Hawking quanta in analogue gravity},
   volume={25},
   ISSN={1878-1535},
   url={http://dx.doi.org/10.5802/crphys.239},
   DOI={10.5802/crphys.239},
   number={S2},
   journal={Comptes Rendus. Physique},
   publisher={MathDoc/Centre Mersenne},
   author={Del Porro, Francesco and Liberati, Stefano and Schneider, Marc},
   year={2025},
   month=Feb, pages={1–27} }

@article{Nguyen_2015,
   title={Acoustic Black Hole in a Stationary Hydrodynamic Flow of Microcavity Polaritons},
   volume={114},
   ISSN={1079-7114},
   url={http://dx.doi.org/10.1103/PhysRevLett.114.036402},
   DOI={10.1103/physrevlett.114.036402},
   number={3},
   journal={Physical Review Letters},
   publisher={American Physical Society (APS)},
   author={Nguyen, H. S. and Gerace, D. and Carusotto, I. and Sanvitto, D. and Galopin, E. and Lemaître, A. and Sagnes, I. and Bloch, J. and Amo, A.},
   year={2015},
   month=Jan }

@article{Williamson,
  title={On the Algebraic Problem Concerning the Normal Forms of Linear Dynamical Systems},
  author={Jack Williamson},
  journal={American Journal of Mathematics},
  year={1936},
  volume={58},
  pages={141},
  url={https://api.semanticscholar.org/CorpusID:124759285}
}

@article{Carusotto_numerical,
   title={Numerical observation of Hawking radiation from acoustic black holes in atomic Bose–Einstein condensates},
   volume={10},
   ISSN={1367-2630},
   url={http://dx.doi.org/10.1088/1367-2630/10/10/103001},
   DOI={10.1088/1367-2630/10/10/103001},
   number={10},
   journal={New Journal of Physics},
   publisher={IOP Publishing},
   author={Carusotto, Iacopo and Fagnocchi, Serena and Recati, Alessio and Balbinot, Roberto and Fabbri, Alessandro},
   year={2008},
   month=Oct, pages={103001} }

\end{document}